\documentclass[fleqn,usenatbib]{mnras}

\usepackage[T1]{fontenc}

\DeclareRobustCommand{\VAN}[3]{#2}
\let\VANthebibliography\thebibliography
\def\thebibliography{\DeclareRobustCommand{\VAN}[3]{##3}\VANthebibliography}

\usepackage{graphicx}	
\usepackage{amsmath}	
\usepackage{amssymb}	
\usepackage{xcolor}

\usepackage{newtxtext,newtxmath}

\title[Non-standard stellar evolution products in NGC\,6791]{Asteroseismology of overmassive, undermassive, and potential past members of the open cluster NGC\,6791.}

\author[K. Brogaard et al.]{K. Brogaard,$^{1,2}$\thanks{E-mail: kfb@phys.au.dk}
T. Arentoft,$^{1}$
J. Jessen-Hansen$^{1}$ and A. Miglio$^{3,4,5}$
\\
$^{1}$Stellar Astrophysics Centre, Department of Physics and Astronomy, Aarhus University, Ny Munkegade 120, DK-8000 Aarhus C, Denmark\\
$^{2}$Astronomical Observatory, Institute of Theoretical Physics and Astronomy, Vilnius University, Saul\.{e}tekio av. 3, 10257 Vilnius, Lithuania\\
$^{3}$Dipartimento di Fisica e Astronomia, Universit\`{a} degli Studi di Bologna, Via Gobetti 93/2, I-40129 Bologna, Italy \\
$^{4}$INAF – Osservatorio di Astrofisica e Scienza dello Spazio di Bologna, Via Gobetti 93/3, I-40129 Bologna, Italy\\
$^{5}$School of Physics and Astronomy, University of Birmingham, Edgbaston, B15 2TT, UK   
}

\date{Accepted XXX. Received YYY; in original form ZZZ}

\pubyear{2021}

\begin{document}
\label{firstpage}
\pagerange{\pageref{firstpage}--\pageref{lastpage}}
\maketitle

\begin{abstract}
We perform an asteroseismic investigation of giant stars in the field of NGC\,6791 with previous indications of atypical evolution. 
The analysis makes use of observations from \textit{Kepler} and Gaia in combination with ground-based photometry, a literature radial-velocity study, and measurements of eclipsing binaries in the cluster. We derive mass, radius, effective temperature, evolutionary stage and apparent distance modulus of each target. 
Among the investigated cluster giants we find clear evidence of overmassive and undermassive members, and non-members with strong hints of potential past membership.
Our results indicate that about 10\% of the red giants in the cluster have experienced mass-transfer or a merger.
High-resolution high-S/N spectroscopic follow-up could confirm potential past membership of the non-members, and reveal whether certain element abundances might expose the non-standard evolution of overmassive and undermassive stars. If so, field stars of similar type could be identified as what they are, i.e. over- or undermassive stars, and not mistakenly classified as younger or older than they are.
\end{abstract}




\section{Introduction}

The field of asteroseismology of giant stars has matured in recent years, and now allow for estimates of radius, mass, and age of stars from the average asteroseismic measures $\Delta \nu$ and $\nu_{\rm max}$. Though the level of accuracy is not yet fully established \citep{Brogaard2016, Brogaard2018dEBs}, comparisons with independent measures in star clusters and detached eclipsing binaries indicate that accuracy is achieved to within the level of the measurement precision when theoretical corrections to the asteroseismic scaling relations are taken into account.
Building upon this success, the field of Galactic archeology has advanced by making use of asteroseismic age estimates for large numbers of stars observed by CoRoT, \textit{Kepler} and K2, with further potential expected from TESS and PLATO.
Such studies are however not without complications. One challenge is that ages of stars are inferred by evolving a stellar model of the measured mass until radius, $T_{\rm eff}$, and/or luminosity also agrees with observations. The age derived thus assumes that the star evolved as a single star. Many stars are however born as binaries, or even triples or higher multiples, and there is plenty of evidence of stars that result from mergers or mass transfer in such systems. The blue straggler V106 \citep{Brogaard2018} is a specific example in the cluster of the present study. 
In asteroseismic studies of field stars, such stars will appear artificially young, since their relatively large mass results in a young age when interpreted as a single star that evolved in isolation. This was indeed what happend in the studies by \citet{Chiappini2015} and \citet{Martig2015} who found apparently young thick disc stars in the Milky Way. Since then, the more plausible interpretation that the stars were in fact old stars that experienced mass-transfer or a merger, has been preferred by observations \citep[e.g.][]{Jofre2016,Yong2016,Izzard2018,SilvaAguirre2018,Miglio2021}. \citet{Brogaard2016} made a rough estimate of the expected number of such stars through an asteroseismic study of the open cluster NGC\,6819, and this was further detailed by \citet{Handberg2017}. These results are however based on observations of only one open cluster and therefore quite uncertain.

NGC\,6791 is a populous old open cluster and therefore suitable for a study of the relative number of overmassive stars compared to the 'normal' stars that evolved as single. This cluster has already been studied extensively for various purposes. \citet{Platais2011} carried out a proper motion (PM) study and established a set of cluster giants located at unusual positions in the cluster colour-magnitude diagram (CMD). Their interpretation that those stars were horizontal branch stars connecting the red clump stars to the hot horizontal branch stars of the cluster was challenged by \citet{Brogaard2012}, who demonstrated that one particular star, 2-17 \citep{Kinman1965}, was a blue straggler star (BSS), not a hot horizontal branch star.  
Later, \citet{Tofflemire2014} published a radial velocity (RV) study of the cluster and revisited the sample of \citet{Platais2011}. They found that some were non-members, and others were close binaries where a history involving mass-transfer is the likely explanation for their CMD positions. In particular, V106 was later established firmly as a BSS member by a detailed study \citep{Brogaard2018}. Still, a small number of giant stars close to, but significantly different from, the well-defined red clump of NGC\,6791, remained as members in the study by \citet{Tofflemire2014}. 

The location of NGC\,6791 in the field of view of the \textit{Kepler} mission \citep{Borucki2010} allowed asteroseismic studies of the giant members, (e.g. \citealt{Stello2011, Basu2011, Corsaro2012, Miglio2012}). However, the targets were chosen to be close to the cluster sequence in the CMDs, and therefore none of the atypical stars established by \citet{Platais2011} and confirmed as members by \citet{Tofflemire2014} were observed as individual targets. The vast majority of the individual cluster targets observed by \textit{Kepler} are 'normal' cluster stars that evolved as single. Only one star was established as a so-called outlier by \citet{Corsaro2012} and interpreted as an overmassive red giant branch star belonging to NGC\,6791 by \citet{Brogaard2012}.

In this paper, we investigate the remaining sample of atypical stars supplemented by a few stars selected to represent the population of normal cluster stars. Among these, we find clear evidence of overmassive cluster members, and a single undermassive member. We also establish the stellar parameters of the non-members in the \citet{Platais2011} sample, showing them to be old stars. Our investigation makes use of asteroseismology exploiting light curves derived from \textit{Kepler} superstamps \citep{Kuehn2015} combined with ground-based photometry and Gaia parallaxes and proper motions. In Sect. \ref{sec:data} we introduce the targets and detail the observations and data used. We also explain how we produced the light curves from \textit{Kepler} data. We then move on to extract oscillation frequencies and derive asteroseismic parameters in Sect. \ref{sec:astparms} and use those, along with supplementary observations to establish stellar parameters in Sect. \ref{sec:stellarparms}. These are discussed and interpreted in Sect. \ref{sec:discussion}-\ref{sec:non-standard}. We conclude in Sect. \ref{sec:conclusion}. 





\section{Targets and observations}
\label{sec:data}

In this study, we investigated six giant stars in the field of the open cluster NGC\,6791, which were claimed to be unusual horizontal branch cluster members in the study by \citet{Platais2011}. To this sample, we added the asteroseismic outlier KIC\,2437589 from \citet{Corsaro2012}, which was suggested to be an overmassive RGB cluster member by \citet{Brogaard2012}. Finally, we also included four giant stars that were meant to serve as a reference representing normal single cluster giants.

The targets were all observed by the \textit{Kepler} mission, but not all as single targets. Five targets, KIC\,2436543, KIC\,2437209, KIC\,2437267, KIC\,2438100 and KIC\,2438139
were only observed as part of larger regions, known as superstamps, covering the central parts of NGC\,6791. KIC2570652 is only observed on a superstamp for half of the observing quarters. For this target, the superstamps were supplemented by target-pixel-files.
We used the python package Lightkurve \citep{LCcollab2018} to define apertures and extract light curves of the targets from either superstamps or target-pixel-files. For one target, KIC2707478, which was not observed on a superstamp, we used the available SAPFLUX light curves from quarters 10-17. 
Individual quarters were combined and the KASOC filter \citep{Handberg2014} was applied to the light curves with a 30 day long, and a 0.5 day short, time scales.

\section{Asteroseismic parameters}
\label{sec:astparms}

\begin{figure*}
	\includegraphics[width=15cm]{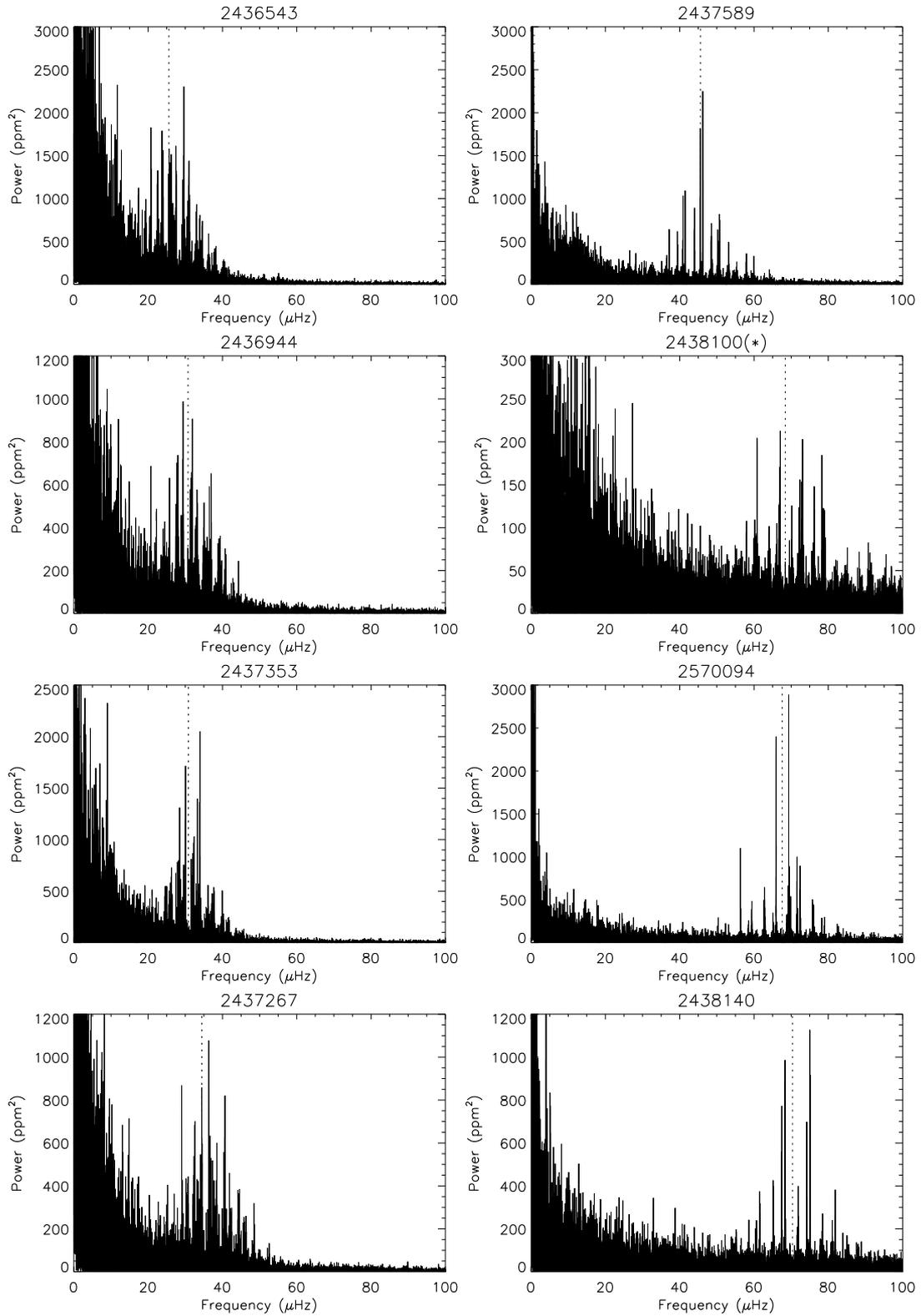}
    \caption{Power spectra of the eight stars classified as cluster members below. KIC\,2438100 has been marked with a (*) as its membership status is uncertain. The oscillations are clearly visible for all eight stars, and the frequency of maximum power, $\nu_{\rm max}$, is for each star indicated by a vertical dotted line.}
    \label{fig:power}
\end{figure*}

\begin{figure*}
	\includegraphics[width=15cm]{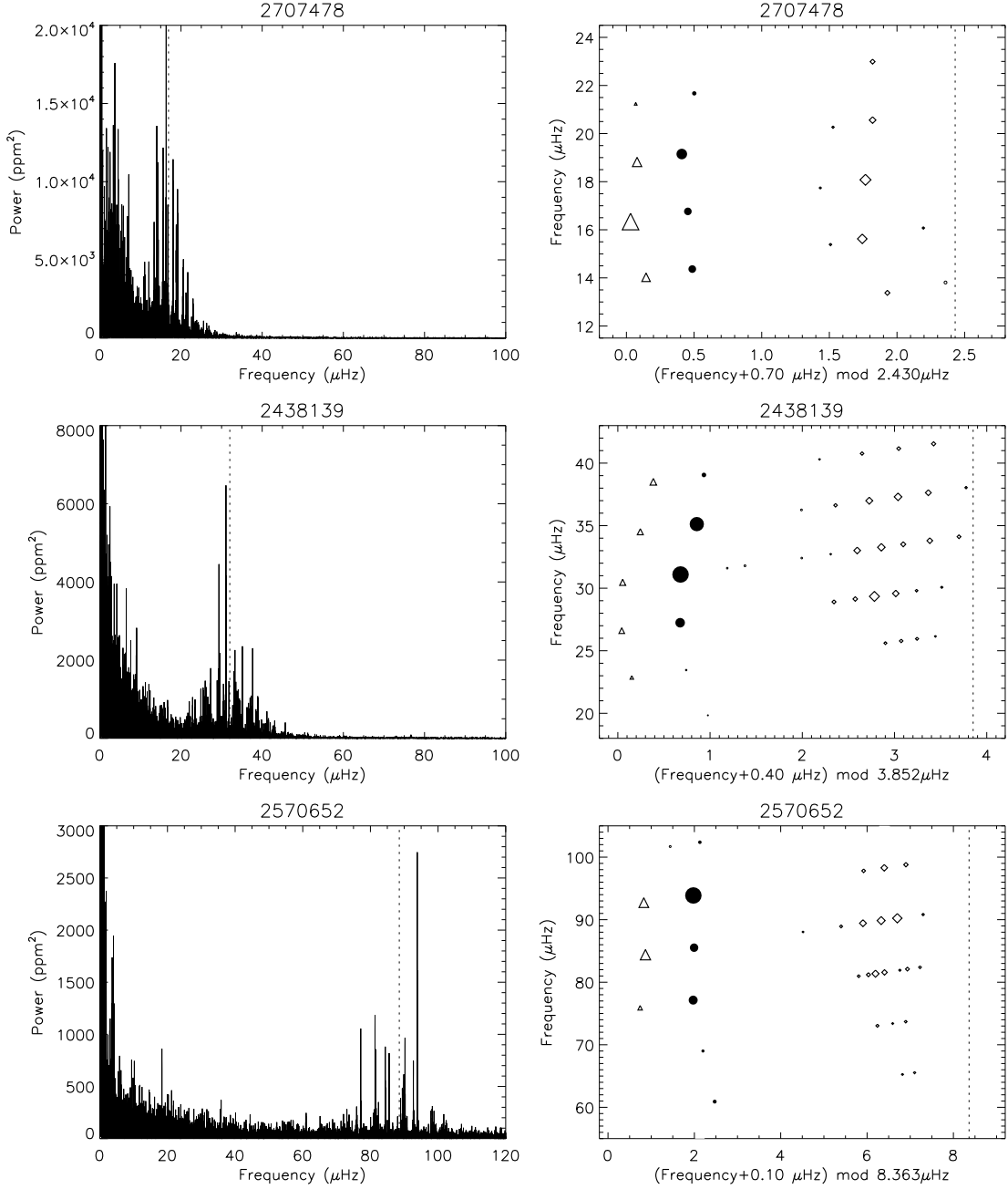}
    \caption{Power spectra (left) and {\'e}chelle diagrams (right) for the three stars, which are classified as non-members below. The vertical dotted lines indicate $\nu_{\rm max}$ in the power spectra and $\Delta\nu$ in the {\'e}chelle diagrams. We have applied a constant shift to the frequencies for constructing the x-axis in the {\'e}chelle diagrams, in order to align the $\ell=0$ modes (filled circles) and $\ell=2$ modes (triangles) on the left side of the diagrams. Modes with $\ell=1$ are shown as diamonds, while modes with uncertain classification are shown as open circles.}
    \label{fig:nonmembersiesmo}
\end{figure*}

In the asteroseismic analysis of the \textit{Kepler} light curves we followed the methods described in \citet{Arentoft17}, with 
only a slight modification introduced in \citet{Arentoft19}. The methods are described in
detail in those papers, and are therefore only briefly recounted here. The analysis is based on the power spectra shown for the cluster members in Fig.\,\ref{fig:power} and, for the non-members, in the left column of Fig.\,\ref{fig:nonmembersiesmo}. 

We started by determining the global oscillation parameters $\nu_{\rm max}$ and $\Delta \nu_{\rm ps}$, which, respectively, is the frequency of maximum power and the large frequency spacing of modes of consecutive radial order, determined as an average value based on the entire oscillation spectrum. The latter value is subsequently refined by investigating the radial oscillation modes (modes with $\ell = 0$) alone. We determined $\nu_{\rm max}$ by performing combined fits of a stellar background and a Gaussian-shaped oscillation envelope to the individual power spectra \citep{Handberg2017,Arentoft17}. The results for $\nu_{\rm max}$ are listed in Table\,\ref{tab:seis} and marked with dotted vertical lines in Fig.\,\ref{fig:power} and in the left column of Fig.\,\ref{fig:nonmembersiesmo}. As in \citet{Arentoft17}, the uncertainty was estimated by splitting each time series in two, performing the same fit to the resulting power spectra, and taking the uncertainty as the largest difference between the results from the fit to the full time-series and to the two half series, divided by $\sqrt{2}$. The resulting uncertainties are rather low, which indicates that our analysis is quite robust. 
The mean large frequency separation, $\Delta \nu_{\rm ps}$, was determined for each star using several methods; we used autocorrelation of the part of the spectrum where the oscillation signal is found, we used the method described in \citet{JCD08} and applied in \citet{Arentoft17} to similar data, where the part of the power spectrum which contains the oscillation signal is cut up in sections of $\Delta \nu / 2$ and stacked for a range of trial $\Delta \nu$-values, and finally a slightly modified version of the latter applied to $\epsilon$\,Tau by \citet{Arentoft19}, where the spectrum is cut up in sections of $\Delta \nu$ instead of $\Delta \nu / 2$, which is more suitable for red giants, see \citet{Arentoft19}. When the correct $\Delta\nu$ is used, the regularly spaced oscillation modes will add up and create a strong signal, see \citet{JCD08} for details. It is the results from the last of these methods which are quoted along with $\nu_{\rm max}$ in Table\,\ref{tab:seis}, however all three methods agreed well for all eleven stars. The uncertainty was again estimated for each star by comparing the results of applying the method to the power spectrum of the full series to those of the two half series. 

\begin{figure*}
	\includegraphics[width=15cm]{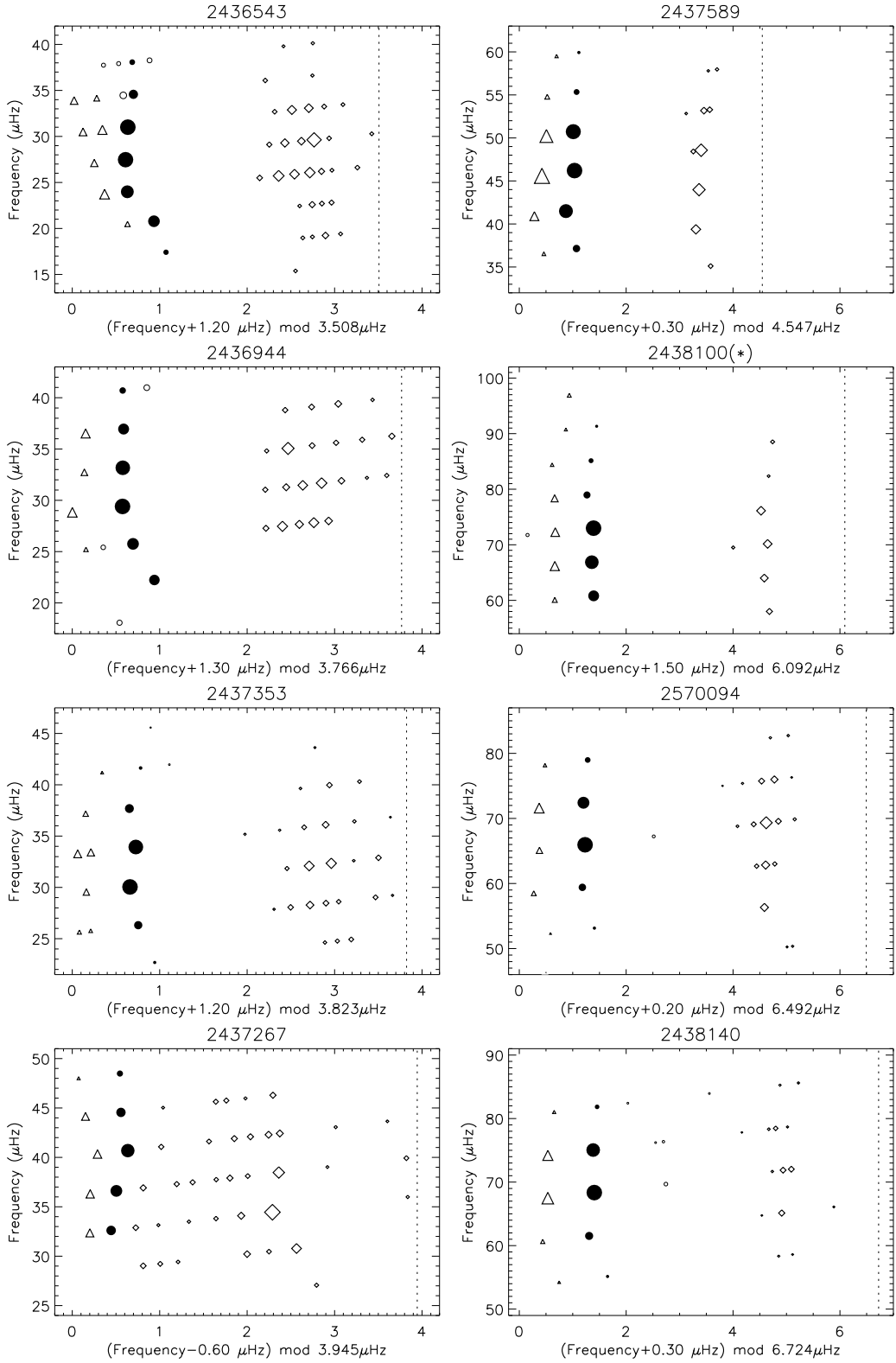}
    \caption{{\'E}chelle diagrams for the eight stars classified as members. KIC\,2438100 is again marked with a (*) due to its uncertain membership status, and we have again applied a constant shift to construct the x-axis. The legends are as in Fig.\,\ref{fig:nonmembersiesmo}. We note that some of the modes classified as $\ell=1$ may be $\ell=3$ and, as was the case in NGC\,6811 \citep{Arentoft17}, some of the $\ell=2$ modes are split in two, which may be due to period spacing for $\ell=2$. Most of the stars show rich mode-structure for $\ell=1$, which allows us to determine the observed period spacing, $\Delta P_{\rm obs}$ and, in some cases, the asymptotic period spacing, $\Delta P$.}
    \label{fig:echelle}
\end{figure*}

We then determined individual frequencies with uncertainties and signal-to-noise ratios (S/N) following closely the method described in detail in \citet{Arentoft17}. For the eleven stars, we determined between 21 and 49 frequencies with S/N up to typically $\sim$20, with two stars having a bit lower S/N-values (KIC\,2436944 and KIC\,2438100, up to $\sim$10) and one star having a bit higher values (KIC\,2437353, up to $\sim$27). We generally only include modes if they have a S/N-value above 3.5, however in some cases we include a few modes with lower S/N if they fit into the expected mode structure for the solarlike oscillations, as in \citet{Arentoft17}. We then used the large frequency separation from the analysis above, $\Delta \nu_{\rm ps}$, to plot and identify the detected modes in {\'e}chelle diagrams, i.e., to assign $\ell$- and, for $\ell=0,2$, also $n$-values to the modes. As illustrated in the rightmost panels of Fig.\,\ref{fig:nonmembersiesmo} and in Fig.\,\ref{fig:echelle}, we applied a constant shift to the x-axis, to make modes of $\ell=0,2$ (filled circles and triangles, respectively) line up on the left in the diagrams. The remaining modes, shown as diamonds, were identified as $\ell=1$, however some of the modes close to the $\ell=0$ ridges may be $\ell=3$. We do not have the means to separate $\ell=3$ modes from $\ell=1$, so we refrain from identifying $\ell=3$ modes, but just note that they may be present. A few modes for some of the stars have uncertain identification, these are shown as open circles in the diagrams. 

With the mode identification in hand, we can determine the rest of the seismic parameters which are listed in Table\,\ref{tab:seis}. We refined our value of $\Delta\nu$ by performing an uncertainty-weighted linear fit to the $\ell=0$ modes, which gave us the values listed as $\Delta\nu_{\ rm 0}$ in Table\,\ref{tab:seis}, along with the $\epsilon$-values for each star, see \citet{Arentoft17} for details. It is this value for the large frequency separation, $\Delta\nu_{\rm 0}$, we use in the analysis described in the sections below. The small frequency separation between modes of $\ell=0,2$ was determined as well, as was the large frequency separation and $\epsilon$ based on only the three central modes closest to $\nu_{\rm max}$, $\Delta\nu_{\rm c}$ and $\epsilon_{\rm c}$, in order to compare to \citet{Kallinger12}, who used these parameters to discriminate between red giants of different evolutionary status, see their Fig.\,4. Some of the stars display split $\ell=2$ modes, which was also the case in NGC\,6811 \citep{Arentoft17}. In those cases, we used the mean value of the split $\ell=2$ modes to determine $\delta\nu_{\rm 02}$. We finally determined the observed period spacing $\Delta P_{\rm obs}$ for the $\ell=1$ modes, and in some cases also the asymptotic period spacing, $\Delta P$, using the same procedures as in \citet{Arentoft17}. $\Delta P_{\rm obs}$ can be used to discriminate between hydrogen-shell-burning and helium-core-burning giants, the former having observed period-spacing values around 50 seconds, the latter in the range 100--300 seconds \citep{Bedding11}. We can immediately see from the values listed in Table\,\ref{tab:seis} that five of our stars can be classified as helium-core-burning giants, while three stars with $\Delta P_{\rm obs}\sim50$ are hydrogen-shell-burning giants. We were not able to determine $\Delta P_{\rm obs}$ for three of the eleven stars. In this paper we refer to the hydrogen-shell-burning giants as red giant branch (RGB) stars and helium-core-burning as red clump (RC) stars. 

We compared our results for $\Delta\nu$ and $\nu_{\rm max}$ to those of \citet{Bossini2020}, with whom we have five stars in common; KIC\,2436944, KIC\,2437353, KIC\,2570094, KIC\,2438140 and KIC\,2437589. They do not quote uncertainties for their values, but if these are similar to ours, the results agrees well within 1\,$\sigma$, or in some cases just slightly more than 1$\sigma$. Furthermore, based on the period spacing they conclude that KIC\,2436944 and KIC\,2437353 are RC stars while KIC\,2570094 and KIC\,2438140 are RGB, in agreement with our results. Like us, \citet{Bossini2020} did not find a period spacing for KIC\,2437589. They list the evolutionary status as unclear, but probably RGB, which is also the case when we compare our measurements of $\Delta\nu_{\rm c}$, $\epsilon_{\rm c}$ and $\delta\nu_{\rm 02} / \Delta \nu$ to Fig.\,4 of \citet{Kallinger12}: KIC\,2437589 is in the upper and lower panels of this figure placed among the hydrogen-shell-burning RGB stars, but the group of red clump stars lies within 1$\sigma$ in both panels, making a classification based on these diagrams ambiguous. The same is the case for the five stars in Table\,\ref{tab:seis} for which $\Delta P_{\rm obs}$ points to a RC-classification. These stars lie among the RC-stars in Fig.\,4 of \citet{Kallinger12}, but with the RGB-stars within 1$\sigma$. They are, however, RC stars based on their observed period spacing. The three RGB-stars in Table\,\ref{tab:seis} with $\Delta P_{\rm obs}$ close to 50\,s can be classified as RGB-stars from the diagrams of \citet{Kallinger12} as well. We have two more stars in Table\,\ref{tab:seis} for which we could not determine a period spacing. For KIC\,2707478, the positions in the diagrams of Fig.\,4 of \citet{Kallinger12} suggest that it is a RGB-star (but note that the possibility of AGB is not considered in these diagrams), while the same diagrams suggest that KIC\,2438100 seems to belong to the secondary clump from these criteria, although a RGB-classification cannot be ruled out. The evolutionary status of KIC\,2438100 will be discussed in detail below. 


%
%
%
%

\begin{table*}
\caption{Asteroseismic parameters for eleven giants in NGC\,6791.}
\begin{tabular}{ccccccccccc}
\hline
\noalign{\smallskip}
KIC & $\Delta\nu_{\rm ps}$ ($\mu$Hz) & $\Delta\nu_{\rm 0}$ ($\mu$Hz) & $\delta_{\rm 02}$ ($\mu$Hz) & $\epsilon$ & $\nu_{\rm max}$ ($\mu$Hz) & $\Delta$P$_{\rm obs}$ (s) & $\Delta$P (s) & $\Delta\nu_{\rm c}$ ($\mu$Hz) & $\epsilon_{\rm c}$ & $\delta\nu_{\rm 02}$/$\Delta\nu$ \\
\noalign{\smallskip} \hline
2707478 & 2.565$\pm$0.267  & 2.430$\pm$0.022 & 0.365$\pm$0.051 &    0.903$\pm$0.061 &  16.92$\pm$0.24 &     -       &  -   &  2.391$\pm$0.004 & 1.009$\pm$0.009 & 0.150$\pm$0.021 \\
2436543 & 3.508$\pm$0.012  & 3.528$\pm$0.008 & 0.307$\pm$0.049 & 0.793$\pm$0.018 &  25.58$\pm$0.48 & 209$\pm$10 & - & 3.511$\pm$0.014 & 0.830$\pm$0.027 & 0.087$\pm$0.014 \\
2436944 & 3.766$\pm$0.042  & 3.705$\pm$0.024 & 0.496$\pm$0.070 & 0.967$\pm$0.050 &  30.73$\pm$0.13 & 223$\pm$6 & 295 & 3.708$\pm$0.036 & 0.942$\pm$0.068 & 0.134$\pm$0.019 \\
2437353 & 3.823$\pm$0.021  & 3.799$\pm$0.022 & 0.529$\pm$0.052 & 0.921$\pm$0.043 &  30.82$\pm$0.25 & 236$\pm$9  &  - & 3.809$\pm$0.046 & 0.901$\pm$0.087 & 0.139$\pm$0.014  \\
2438139 & 3.852$\pm$0.032  & 3.909$\pm$0.026 & 0.625$\pm$0.011 & 0.981$\pm$0.047 &  32.03$\pm$0.10 & 254$\pm$6 & 319   & 3.942$\pm$0.050 & 0.903$\pm$0.090 & 0.160$\pm$0.003  \\
2437267 & 3.945$\pm$0.088  & 3.970$\pm$0.021 & 0.350$\pm$0.053 & 1.223$\pm$0.049 &  34.44$\pm$0.50 & 288$\pm$12           &  - & 4.040$\pm$0.020 & 1.067$\pm$0.041 & 0.088$\pm$0.013  \\
2437589 & 4.547$\pm$0.064  & 4.570$\pm$0.019 & 0.567$\pm$0.057 & 1.107$\pm$0.041 &  45.58$\pm$0.30 &     -     & - & 4.616$\pm$0.053 & 0.997$\pm$0.105 & 0.124$\pm$0.012 \\
2438100 & 6.092$\pm$0.112  & 6.095$\pm$0.016 & 0.674$\pm$0.067 & 0.972$\pm$0.031 &  68.41$\pm$0.21 &    -      & - & 6.091$\pm$0.019 & 0.982$\pm$0.032 & 0.111$\pm$0.011 \\
2570094 & 6.492$\pm$0.018  & 6.469$\pm$0.030 & 0.842$\pm$0.042 & 1.200$\pm$0.043 &  67.59$\pm$0.21 & 52$\pm$3  & 73 & 6.502$\pm$0.023 & 1.139$\pm$0.033 & 0.130$\pm$0.033 \\
2438140 & 6.724$\pm$0.010  & 6.692$\pm$0.044 & 0.859$\pm$0.037 & 1.218$\pm$0.060 &  70.36$\pm$0.72 & 52$\pm9$  & -  & 6.762$\pm$0.034 & 1.101$\pm$0.046 & 0.128$\pm$0.006 \\
2570652 & 8.363$\pm$0.010  & 8.296$\pm$0.039 & 1.168$\pm$0.054 & 1.321$\pm$0.042 &  88.54$\pm$0.63 & 49$\pm$4  & -  & 8.365$\pm$0.011 & 1.222$\pm$0.012 & 0.141$\pm$0.007 \\
\hline
\noalign{\smallskip}
\end{tabular}
\label{tab:seis}
\end{table*}

\section{Stellar parameters}
\label{sec:stellarparms}

To infer stellar parameters, we combine the asteroseismic parameters with the following information: 1) Optical photometry from \citet{Stetson2003} and \citet{Brogaard2012}, 2MASS photometry \citep{Skrutskie2006}, and Gaia EDR3 photometry \citep{Riello2020}. 2) Gaia Parallaxes EDR3 and proper motions \citep{Gaiamission2020,Lindegren2021}. 3) Radial velocity information from \citet{Tofflemire2014}. 

We used the bolometric correction tables of \citet{Casagrande2014,Casagrande2018} to convert observed colours to $T_{\rm eff}$.

Literature reddening values $E(B-V)$ for NGC\,6791 vary from 0.10 to 0.16 \citep[][and references therein]{An2015}. To determine our reddening value, we first required agreement between the spectroscopic and photometric $T_{\rm eff}$ of the components of the eclipsing binary V18 \citep{Brogaard2011} where the latter was determined using the bolometric correction tables of \citet{Casagrande2014} and the observed $B-V$ component colours. This yielded $E(B-V)=0.178$ for an adopted [Fe/H]=$+0.29$ as measured by \citet{Brogaard2011} for the binaries on the upper main sequence. Since this reddening value is larger than the literature values, and since the spectroscopic $T_{\rm eff}$ measurements have an uncertainty of$\pm 125$\,K, we adopted a value $E(B-V)$=0.16, which is consistent with both V18 and the upper end of the the literature values. $T_{\rm eff}$ of the giant stars were then adjusted independently for $V-K_s$ and $G_{\rm BP}-G_{\rm RP}$ until the bolometric corrections matched the observed colour at [Fe/H]=$+0.35$, which takes into account likely diffusion effects on [Fe/H] for the cluster giants relative to [Fe/H]=$+0.29$ at the turn-off \citep{Brogaard2012}. An iteration with the next step was needed to include the correct log$g$ in the transformations.

Since our adopted reddening is on the high side of the mean literature values, we investigated whether any of our target stars had a spectroscopic $T_{\rm eff}$ estimate in the literature that we could compare to. Perhaps surprisingly, we found that all optical spectroscopic investigations of red giant stars in NGC6791 relied on photometric $T_{\rm eff}$ estimates where some reddening value was adopted without strong arguments. \citet{Villanova2018} mention that they do attempt to derive purely spectroscopic values, but that this turned out to be less precise, causing them to revert to the use of photometric $T_{\rm eff}$ estimates. The only exception is the APOGEE infrared spectroscopic survey \citep{Majewski2017}, which however rely on their spectroscopic temperatures being calibrated onto a photometric scale \citep{Holtzman2018,Jonsson2020}. Two of our targets have APOGEE-2 DR16 \citep{Jonsson2020} $T_{\rm eff}$ values, which we show in Table~\ref{tab:starpars} for comparison to our derived values. As seen, our values based on Gaia colours are about 60 K and 10 K hotter than DR16, respectively, for the two stars, suggesting that our reddening value is not far off. We found that $\pm 0.01$ to $E(B-V)$ corresponds to about $\pm 20 K$. We also show $T_{\rm eff}$, [Fe/H] and log$g$ from \citet{Villanova2018} for two overlapping targets in Table ~\ref{tab:starpars}. Their $T_{\rm eff}$ values are about 140 K cooler than our coolest estimates (from $G_{\rm BP}-G_{\rm RP}$), which according to our approximate conversion above would give $E(B-V)=0.09$, lower than any value in the literature for NGC\,6791, and 0.04 mag lower than adopted in their analysis. The main explanation is that they used different colour-temperature relations. We tried their colour combinations with our relations for KIC\,2436543 and found a value close to 4500\,K,  100\,K hotter than their value, considering that they also applied an additional $-25$\,K as suggested in their spectroscopic analysis. Two things, in addition to the low reddening, suggest that their temperatures are too cool. First, their metallicity is the same as derived for turn-off stars in the cluster by \citet{Brogaard2011} suggesting either that element diffusion does not take place in the cluster or that $T_{\rm eff}$ should be higher. Second, their log$g$ values are 0.3 dex too low compared to our asteroseismic values, also indicating that the true effective temperature values are larger. Summarising, it appears that both photometric and spectroscopic measurements support that our $T_{\rm eff}$ values are likely close to the true values. The difference between our estimates from $V-K_s$ and $G_{\rm BP}-G_{\rm RP}$ suggest an uncertainty of about 100\,K or less in general, but larger in a few cases. We have not averaged the $T_{\rm eff}$ estimates from the two colours, but kept both to examine the effects of adopting either set.

With $T_{\rm eff}$ in hand, we calculated masses and radii for the giants using the asterosesimic scaling relations. 
These are based on the asteroseismic parameters $\nu_{\mathrm{max}}$, the frequency of maximum power, and $\Delta \nu$, the large frequency spacing between modes of the same radial degree. $\Delta \nu$ is known to scale approximately with the mean density of a star \citep{Ulrich1986} while $\nu_{\mathrm{max}}$ scales approximately with the acoustic cut-off frequency of the atmosphere, which is related to surface gravity and effective temperature \citep{Kjeldsen1995,Belkacem2011}. The relations can be written as



\begin{eqnarray}\label{eq:01}
\frac{\Delta \nu}{\Delta \nu _{\odot}} & = & f_{\Delta \nu}\left(\frac{\rho}{\rho_{\odot}}\right)^{1/2},\\
\label{eq:02}
\frac{\nu _{\mathrm{max}}}{\nu _{\mathrm{max,}\odot}} & = & f_{\nu _{\mathrm{max}}} \frac{g}{g_{\odot}}\left(\frac{T_{\mathrm{eff}}}{T_{\mathrm{eff,}\odot}}\right)^{-1/2},
\end{eqnarray}

where $\rho$, $g$, and $T_{\rm eff}$ are the mean density, surface gravity, and effective temperature, and we have adopted the notation of \citet{Sharma2016} and \citet{Brogaard2018dEBs} that includes the correction functions $f_{\Delta \nu}$ and $f_{\nu _{\mathrm{max}}}$. We adopt the solar reference values from \citet{Handberg2017}, $\Delta \nu _{\odot} = 134.9\, \mu$Hz and $\nu _{\mathrm{max,}\odot} = 3090\, \mu$Hz. 
By rearranging, expressions for the radius and mass can be obtained: 

\begin{eqnarray}\label{eq:03}
\frac{R}{\mathrm{R}_\odot} & = & \left(\frac{\nu _{\mathrm{max}}}{f_{\nu _{\mathrm{max}}}\nu _{\mathrm{max,}\odot}}\right) \left(\frac{\Delta \nu}{f_{\Delta \nu}\Delta \nu _{\odot}}\right)^{-2} \left(\frac{T_{\mathrm{eff}}}{T_{\mathrm{eff,}\odot}}\right)^{1/2},\\ 
\label{eq:04}
\frac{M}{\mathrm{M}_\odot} & = & \left(\frac{\nu _{\mathrm{max}}}{f_{\nu _{\mathrm{max}}}\nu _{\mathrm{max,}\odot}}\right)^3 \left(\frac{\Delta \nu}{f_{\Delta \nu}\Delta \nu _{\odot}}\right)^{-4} \left(\frac{T_{\mathrm{eff}}}{T_{\mathrm{eff,}\odot}}\right)^{3/2}.
\end{eqnarray}

Some empirical tests of these equations have been performed, e.g. \citet{Brogaard2012, Miglio2012, Handberg2017}. A much larger effort is still needed to establish the obtainable accuracy in general. However, it appears that using a correction $f_{\Delta \nu}$ calculated from models \citep{Rodrigues2017}, and assuming $f_{\nu _{\mathrm{max}}}=1$ reproduces masses and radii for red giants at a level of a few percent \citep{Brogaard2018dEBs}. We therefore adopted that procedure here. First, calculations were done without corrections to $\Delta \nu$, to obtain a mass estimate, and then we iterated using the theoretical corrections to $\Delta \nu$ from \citet{Rodrigues2017}. To obtain $f_{\Delta \nu}$, the evolutionary status is needed, for which employed our $\Delta P_{\rm obs}$ measurements. For the three stars where we could not determine $\Delta P_{\rm obs}$ we relied on alternative indications, as detailed later. 

Assuming that the targets are cluster members, and employing Gaia EDR3 parallaxes together with the photometry and $T_{\rm eff}$ estimates allowed us to also use the asteroseismic scaling for mass in three additional forms as first done by \citet{Miglio2012}:

\begin{eqnarray}
\label{eq:05}
\frac{M}{\mathrm{M}_\odot} & = & \left(\frac{\Delta \nu}{f_{\Delta \nu}\Delta \nu _{\odot}}\right)^{2} \left(\frac{L}{L_{\odot}}\right)^{3/2} \left(\frac{T_{\mathrm{eff}}}{T_{\mathrm{eff,}\odot}}\right)^{-6}\\
\frac{M}{\mathrm{M}_\odot} & = & \left(\frac{\nu _{\mathrm{max}}}{f_{\nu _{\mathrm{max}}}\nu _{\mathrm{max,}\odot}}\right) \left(\frac{L}{L_{\odot}}\right) \left(\frac{T_{\mathrm{eff}}}{T_{\mathrm{eff,}\odot}}\right)^{-7/2}\\
\frac{M}{\mathrm{M}_\odot} & \simeq & \left(\frac{\nu _{\mathrm{max}}}{f_{\nu _{\mathrm{max}}}\nu _{\mathrm{max,}\odot}}\right)^{12/5} \left(\frac{\Delta \nu}{f_{\Delta \nu}\Delta \nu _{\odot}}\right)^{-14/5} \left(\frac{L}{L_{\odot}}\right)^{3/10}
\end{eqnarray}

A complication with this is the well-known issue of a zero-point offset in the Gaia DR2 and EDR3 parallax values \citep[e.g.][]{Lindegren2021,Stassun2021}. For the EDR3 parallaxes, we give both the catalog values and those corrected according to \citet{Lindegren2021}, and we adopt the latter. As indicated by \citet{Stassun2021} through a study of eclipsing binaries, there could be an additional offset of 15 $\mu$as, although not statistically significant in their study. For all conversions between parallax and distance we used the simple inversion $d=1/\pi$. This procedure is valid in our case since all targets except one have $\sigma_{\pi}/\pi$ < 0.08 \citep[see panel (i) of figure 6 ]{Bailer2018}.
Due to the uncertainty on the exact parallax zero-point,
we decided to also determine our own parallax estimate as part of the analysis. We did that by adjusting the parallax zero-point until the mass scatter from the four mass equations for all members was minimized. Note that this corresponds to requiring the radius to be identical as determined from both Eq. (3) and $L=R^2T_{\rm eff}^4$ with $L$ and $T_{\rm eff}$ derived from observed magnitudes and bolometric corrections.


As an alternative to the Gaia EDR3 parallax, we also repeated the procedure assuming the apparent distance modulus as determined using turn-off eclipsing binaries \citep{Brogaard2011,Brogaard2012}, $(m-M)_V=13.51$ along with our adopted reddening $E(B-V)=0.16$ and the $V,(V-K_s)$ or the $G,(G_{\rm BP}-G_{\rm RP})$ photometry. 

Our procedures resulted in a number of estimates for various properties of the targets. All this information is available in Table \ref{tab:starpars}. The first entries are multiple identification IDs for the targets. Then follows our $T_{\rm eff}$ estimates along with a comparison to values from \citet{Majewski2017} and \citet{Villanova2018} for overlapping targets. After that we display dynamical values, i.e., RV measurements from \citet{Tofflemire2014} and PM values from Gaia DR2 and EDR3. Parallax values from Gaia are shown for DR2 and EDR3, the latter both with and without the correction to the parallax zeropoint suggested by \citet{Lindegren2021}. For the corrected EDR3 parallaxes we also give the uncertainty and $\sigma_{pi}/\pi$
and the true distance modulus values derived by inverting the parallax.
Then follows the asteroseismic properties, starting with evolutionary state and theoretical correction to $\Delta \nu$.
A section gives the asteroseismic mass, radius, luminosity, surface gravity, and a derived parallax based on the asteroseissmic scaling relations in Eqns. (3) and (4). Three separate sections give mass estimates based on Eqns. (4) to (7) and their mean value and RMS with different assumptions for obtaining $L/L_{\odot}$, as explained above. The last line of the table gives our classification of each giant. 

\begin{figure*}
	\includegraphics[width=18cm]{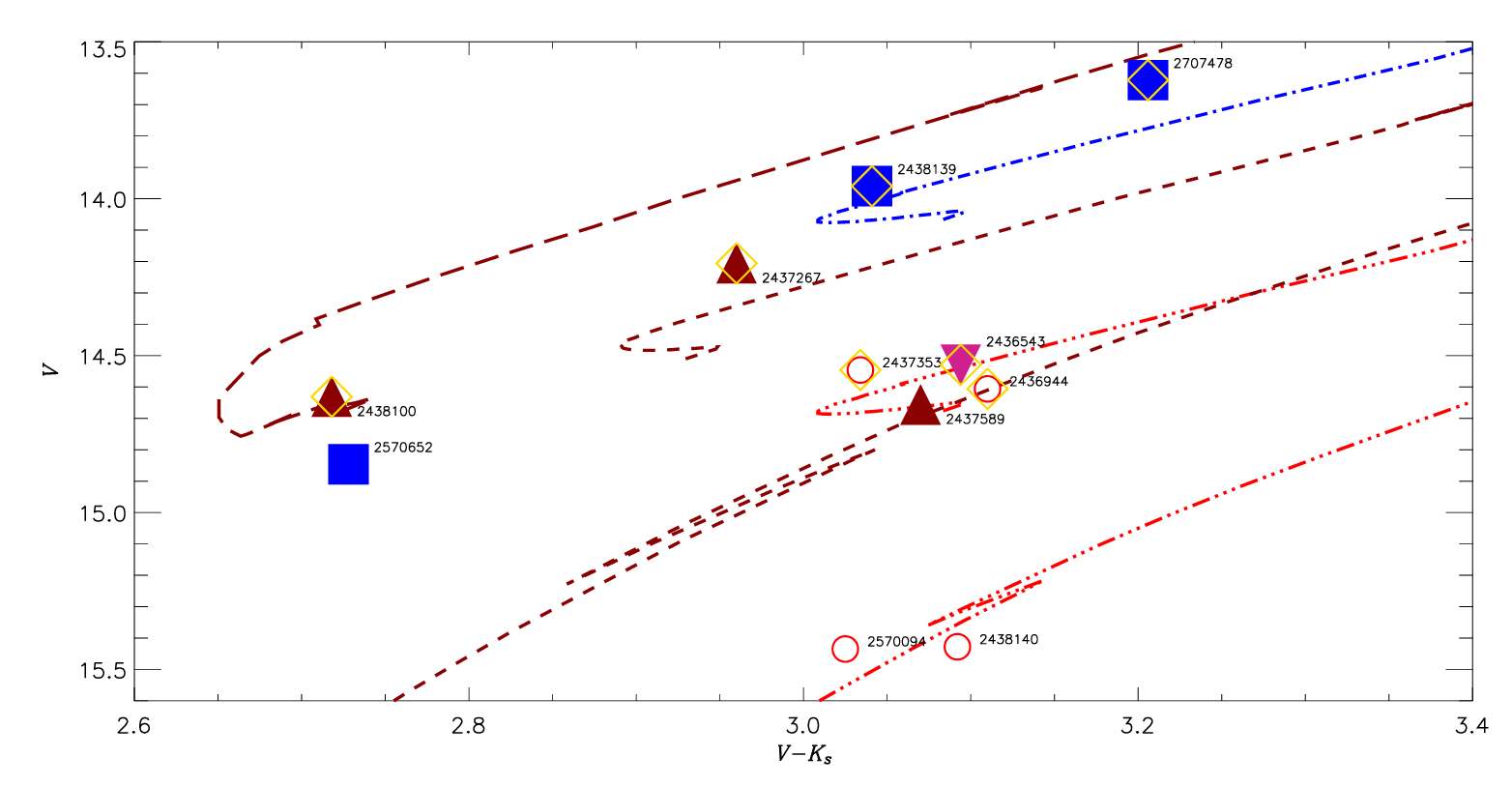}
	\includegraphics[width=18cm]{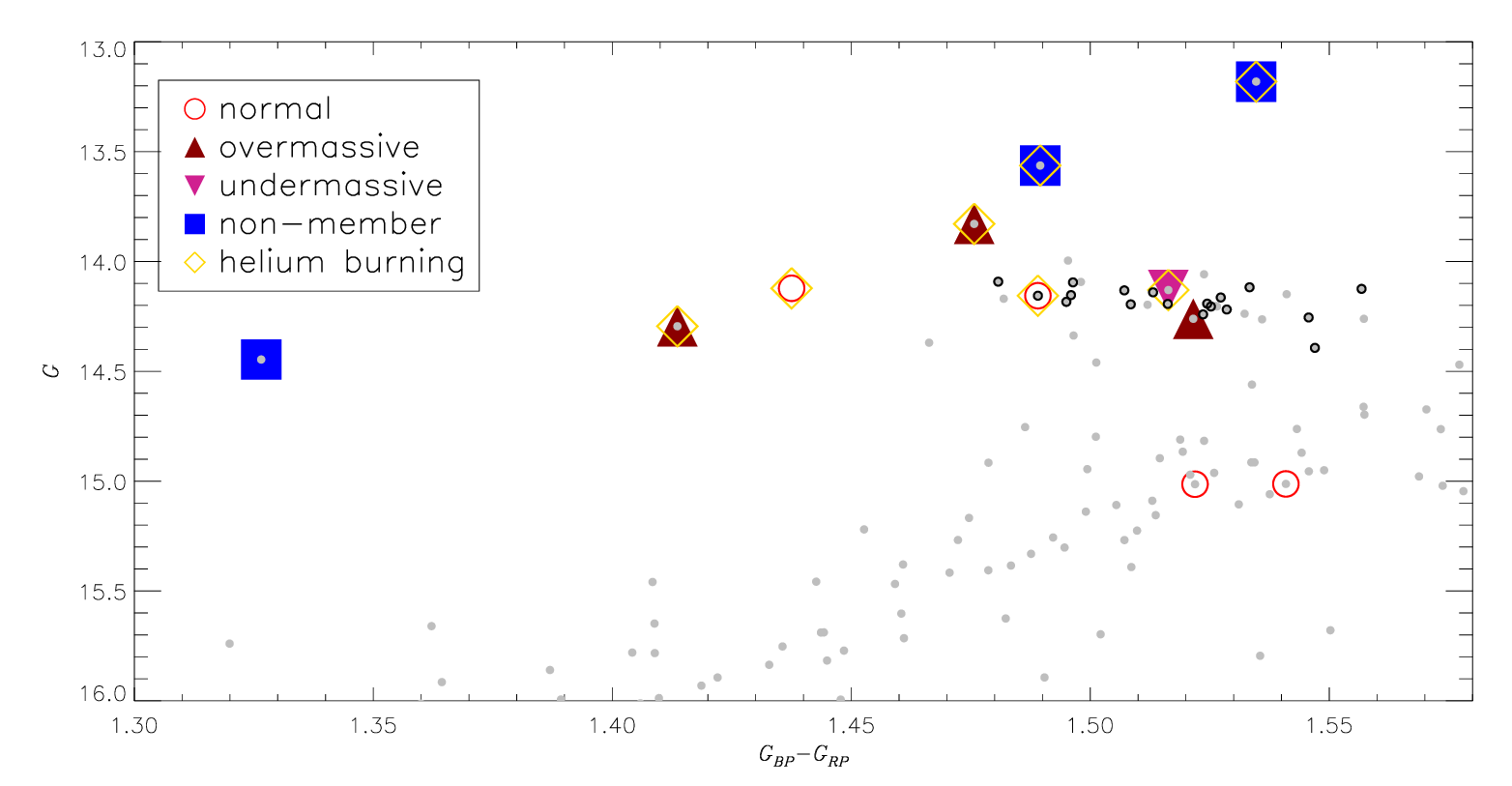}
    \caption{Colour-magnitude diagrams of the giant stars investigated. 
    \textit{Upper panel}: Blue solid squares are non-members. 
    Red circles are members of normal mass. Dark red triangles are overmassive members. Purple downwards pointing triangle is an undermassive cluster member. Yellow diamonds are helium burning stars.
    PARSEC isochrones \citep{Bressan2012} are plotted to allow a comparison to expected locations. The red triple-dot-dashed lines is an 8.5 Gyr, $\rm{[Fe/H]}=+0.35$ isochrone at $(m-M)_V=13.51$ with an RGB mass of 1.12 $M_{\odot}$ and a RC mass of 1.07 $M_{\odot}$. The blue dot-dashed isochrone represents the helium-burning part of the same isochrone, but shifted to the approximate apparent distance modulus of the non-members. The dashed dark red isochrone assumes the same metallicity and apparent distance modulus but has an age of 3.5 Gyr, which corresponds to masses of 1.48 $M_{\odot}$ for both the RGB and RC, more representative of two of the the overmassive members. The long-dashed isochrone represents the helium-burning part for an age of 1.65 Gyr corresponding to a mass of 1.89 $M_{\odot}$, properties which should resemble those of KIC\,2438100. All isochrones have been shifted by the same amount of reddening, $E(B-V)=0.08$ for $E(V-K)=2.72\times E(B-V)$, which is however not the assumed reddening of the stars, due to a well-known issue that stellar models are often not able to reproduce the observed $T_{\rm eff}$s and colours of giants.
    \textit{Lower panel}:Gray filled circles are cluster members according to \citet{Cantat2018} (membership probability $\ge0.6$). Note that our non-members are members according to \citet{Cantat2018}. KIC\,2436944 was not considered by \citet{Cantat2018} because it does not have proper motion values in Gaia DR2. This star has a nearby companion which has likely also affected the Gaia colour, as also indicated by the quite different location in the upper panel.
    Black open circles are RC stars investigated asteroseismically by \citet{Bossini2020}. Other symbols are as described in the upper panel.}
    \label{fig:CMD}
\end{figure*}

\begin{figure*}
	\includegraphics[width=18cm]{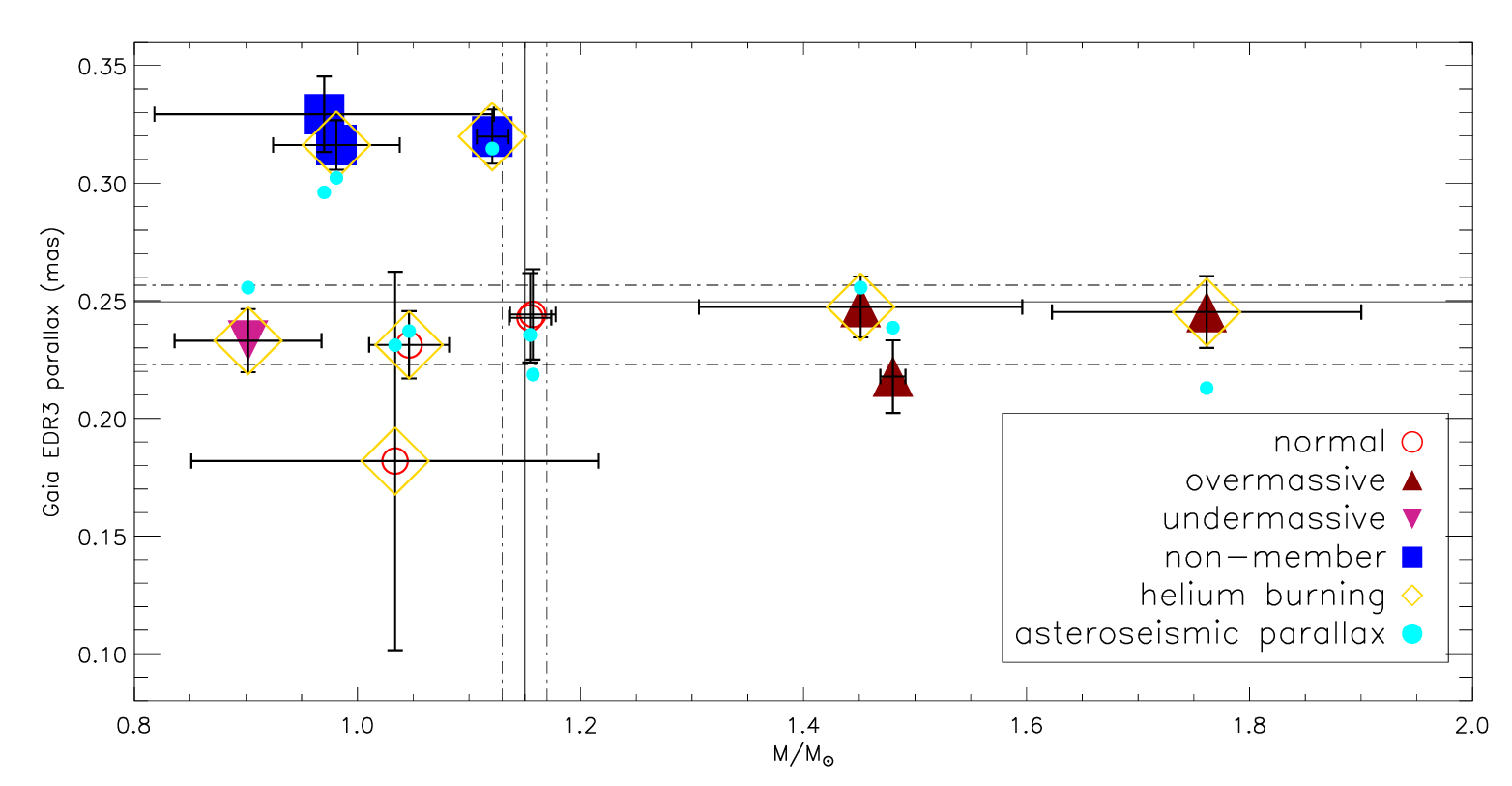}
    \caption{Mass and parallax of the giants. The parallaxes are the Gaia EDR3 parallaxes corrected according to \citet{Lindegren2021}. The masses are derived using the Gaia photometry assuming $(m-M)_V=13.51\pm0.06$ and $E(B-V)=0.16$ for the members and assuming the Gaia EDR3 parallax for the non-members. The mass errorbars are determined from the rms of the mass derived using the four different mass equations.
    The solid lines are estimates of RGB mass and parallax for NGC\,6791 using measurements of eclipsing binary stars \citep{Brogaard2012} and assuming $E(B-V)=0.16$. The dashed-dotted lines mark the corresponding uncertainty range. For the parallax this range includes the possibility that $E(B-V)$ could be as low as 0.10}
    \label{fig:masspar}
\end{figure*}

Colour-magnitude diagrams of the targets are shown in Fig.~\ref{fig:CMD}, where they are compared to PARSEC isochrones \citep{Bressan2012} in the upper panel and cluster members from \citet{Cantat2018} in the lower panel. This is supplemented by Fig.~\ref{fig:masspar}, which shows mass versus parallax to allow visual separation of members from non-members and overmassive and undermassive stars from normal cluster stars. 

As shown e.g. by \citet{Montalban2012}, in models of both RGB and RC stars, and for a given $\langle\Delta\nu\rangle$, the average separation between frequencies of radial and quadrupolar pressure-dominated modes ($\langle d_{02}\rangle$) decreases with increasing stellar mass. Therefore, using $\langle d_{02}\rangle$ as an additional constraint on stellar mass should in principle be possible. However, \citet{Corsaro2012} and \citet{Handberg2017} have shown that the observational trends do not quite follow the theoretical ones. This discrepancy remains for our observations: $\langle d_{02}\rangle/\langle\Delta\nu\rangle$ values are nearly identical for RGB and RC of low mass while theory predicts them to be significantly different (see e.g. fig. 7 of \citealt{Handberg2017}). As discussed in Joergensen et al. 2021, in prep., however, model-predicted small frequency separations may be significantly biased by the so-called surface effects.  This is particularly relevant for RC stars where, depending on the mass, even the most pressure-like $\nu_{n,\ell=2}$ modes may have mode inertias higher than what radial modes would have at the same frequency (see e.g. fig 3 in \citealt{Montalban2012}). This makes $\nu_{n,\ell=2}$  modes less sensitive to surface effects than radial modes, affecting directly the value of the small frequency separation. While this could potentially explain current discrepancies, more investigations are needed. We therefore refer to Joergensen et al. 2021, in prep. for details and postpone the use of $\langle d_{02}\rangle$ as a mass indicator until a more solid theoretical basis is established.

\begin{table*}
\caption{Stellar parameters for eleven giants in the field of NGC\,6791.}
\begin{tabular}{lccccccccccc}
\hline
\noalign{\smallskip}
KIC & 2570652 & 2438100 & 2436944 & 2437267 & 2438139 & 2707478 & 2436543 & 2437353 & 2570094 & 2438140 & 2437589 \\
Stetson ID$^1$ & 14140 & 12596 & 5712 & 7540 & (...) & 2382 & 3369 & 8082 & 9786 & 12836 & 9462 \\
Platais ID$^2$ & 82982 & 60456 & 69976 & 65895 & 69734 & 89107 & (...) & (...) & (...) & (...) & (...) \\
Brogaard ID$^3$ & (...) & 23528 & 15446 & 17605 & (...) & (...) & 12630 & 18243 & 20160 & 23837 & 19773 \\
Tofflemire ID$^4$ & 21016 & 15007 & 23004 & 2001 & 9007 & 7021 & 11007 & 2002 & 10006 & (...) & 16007 \\
Gaia EDR3 ID$^5$ &205129- & 205110- & 205129- & 205129- & 205110- & 205129- & 205128- & 205129- & 205129- & 205110- & 205110- \\  
& 548888- & 510157- & 325550- & 298062- & 578877- & 816894- & 731126- & 291190- & 370218- & 547953- & 513593- \\
& 6975232 & 8545280 & 3779968 & 5570816 & 3383424 & 6719488 & 8703232 & 6087424 & 0105728 & 5690240 & 8207872 \\

\noalign{\smallskip}
\hline
$T_{\rm eff}(V-K_s)^6$ & 4868.52 & 4877.78 & 4547.76 & 4664.17 & 4600.00 & 4477.46 & 4559.85 & 4605.47 & 4612.50 & 4561.36 & 4578.03 \\
$T_{\rm eff}(G_{\rm BP}-G_{\rm RP})$ & 4927.29 & 4719.06 & 4666.26 & 4584.58 & 4556.94 & 4470.92 & 4505.12 & 4557.90 & 4459.39 & 4494.60 & 4495.27 \\
$T_{\rm eff}(\rm A.DR16)^7$ & - & - & - & - & - & 4411 & - & 4547 & - & - & - \\
$T_{\rm eff}\rm{(V2018)^8}$ & - & - & - & - & - & - & 4376 & 4411 & - & - & - \\
log$g\rm{(V2018)^8}$ & - & - & - & - & - & - & 1.99 & 2.04 & - & - & - \\
$\rm [Fe/H] (V2018)^8$ & - & - & - & - & - & - & $+0.28$ & $+0.31$ & - & - & - \\
RV$^5$            & 11.47 & -44.80 & -47.07 & -47.86 & -22.84 & -71.11 & -47.97 & -46.02 & -47.17 & - & -45.990 \\
RV mem-\%$^5$ & 0 & 56 & 96 & 95 & 0 & (0) & 95 & 91 & 96 & - & 91 \\
Gaia DR2 pmra & 0.210 & -0.508 & --- & -0.467 & 0.266 & -0.736 & -0.362 & -0.373 & -0.467 & -0.409 & -0.422 \\
Gaia DR2 pmdec & -2.571 & -2.193 & --- & -2.140 & -2.416 & -2.549 & -2.132 & -1.924 & -2.064 & -2.302 & -2.260 \\
Gaia EDR3 pmra &  0.248 & -0.473 & -0.348 & -0.488 & 0.311 & -0.711 & -0.343 & -0.384 & -0.384 & -0.373 & -0.392 \\
Gaia EDR3 pmdec & -2.546 & -2.247 & -2.359 & -2.149 & -2.366 & -2.686 & -2.247 & -1.871 & -2.174 & -2.327 & -2.182 \\

Gaia DR2 $\pi$ & 0.264854 & 0.157738 & -1.63291 & 0.183306 & 0.283927 & 0.270139 & 0.184648 & 0.188082 & 0.212177 & 0.181706 & 0.176704 \\
Gaia EDR3 $\pi$ & 0.300591 & 0.216126 & 0.152657 & 0.217855 & 0.290067 & 0.286224 & 0.203514 & 0.201824 & 0.214953 & 0.213612 & 0.188245 \\
Gaia 3ZPc $\pi^9$ & 0.329255 & 0.245181 & 0.181879 & 0.247349 & 0.319754 & 0.316184 & 0.233025 & 0.231232 & 0.244178 & 0.242739 & 0.217749 \\
Gaia EDR3 $\sigma_{\pi}$ & 0.0160  &  0.0152 &    0.0804 &   0.0129 &   0.0115  &  0.0105 &
    0.0134  &  0.0143 &   0.0192 &   0.0190 &   0.0154\\

$\sigma_{\pi}/\pi$ & 0.049 & 0.062 & 0.442 & 0.052 & 0.036 & 0.033 & 0.058 & 0.062 & 0.079 & 0.078 & 0.071\\

$(m-M)_0$|$\pi^{10}$ & 12.4123 & 13.0526 & 13.7011 & 13.0335 & 12.4759 & 12.5003 & 13.1630 & 13.1798 & 13.0615 & 13.0743 & 13.3102 \\
RGB/RC$^{11}$ & RGB & RC* & RC & RC & RC & AGB* & RC & RC & RGB & RGB & RGB* \\
$f_{\Delta \nu}^{12}$ & 0.967 & 0.998 & 0.995 & 1.000 & 1.000 & 0.995 & 1.000 & 0.995 & 0.968 & 0.968 & 0.968 \\
\hline
\multicolumn{12}{l}{Assuming $T_{\rm eff}$ from $(V-K_s)$ :}\\
Derived $\pi$ & 0.296049 & 0.212886 & 0.231123 & 0.255549 & 0.314653 & 0.302232 & 0.255574 & 0.237081 & 0.218632 & 0.235490 & 0.238546 \\
$L/L_{\odot}$ & 31.1767 & 37.7583 & 44.9234 & 60.3504 & 77.8045 & 114.153 & 47.0301 & 45.0550 & 21.2136 & 20.9153 & 41.8446 \\
$R/R_{\odot}$ eqn. (3) & 6.50381 & 9.92564 & 11.5810 & 11.5633 & 11.0159 & 14.7084 & 10.7531 & 11.1171 & 7.96418 & 7.70418 & 10.7213 \\
$M/M_{\odot}$ eqn. (4) & 1.11266 & 2.00419 & 1.18343 & 1.33908 & 1.12244 & 1.04288 & 0.850416 & 1.10063 & 1.23972 & 1.20093 & 1.50937 \\
log$g$ & 2.86033 & 2.73894 & 2.38894 & 2.43460 & 2.40178 & 2.12049 & 2.30164 & 2.38511 & 2.72141 & 2.74056 & 2.55204 \\
\hline
\multicolumn{12}{l}{Assuming Gaia EDR3 parallax and $G,(G_{\rm BP}-G_{\rm RP})$ photometry:}\\
$M/M_{\odot}$ eqn. (4) & 1.13287 & 1.90716 & 1.22998 & 1.30495 & 1.10672 & 1.04060 & 0.835151 & 1.08363 & 1.17851 & 1.17466 & 1.46863 \\
$M/M_{\odot}$ eqn. (5) & 0.796485 & 1.68811 & 2.13435 & 1.66981 & 1.13746 & 0.914926 & 1.20495 & 1.26289 & 1.21021 & 1.23106 & 2.24868 \\
$M/M_{\odot}$ eqn. (6) & 0.895732 & 1.75818 & 1.77614 & 1.53807 & 1.12712 & 0.955033 & 1.06635 & 1.20006 & 1.19955 & 1.21196 & 1.95099 \\
$M/M_{\odot}$ eqn. (7) & 1.05579 & 1.86119 & 1.37332 & 1.37091 & 1.11280 & 1.01415 & 0.898680 & 1.11732 & 1.18478 & 1.18573 & 1.59924 \\
$M/M_{\odot}$ mean & 0.970220 & 1.80366 & 1.62845 & 1.47093 & 1.12103 & 0.981177 & 1.00128 & 1.16597 & 1.19326 & 1.20085 & 1.81689 \\
$\sigma_{M/M_{\odot}}$ & 0.152 & 0.0990 & 0.408 & 0.164 & 0.0139 & 0.0568 & 0.167 & 0.0810 & 0.0143 & 0.0255 & 0.352 \\
\hline
\multicolumn{12}{l}{Assuming $(m-M)_V=13.51, E(B-V)=0.16$ and $V,(V-K_s)$ photometry:}\\
$M/M_{\odot}$ eqn. (4) & 1.11266 & 2.00419 & 1.18343 & 1.33908 & 1.12244 & 1.04288 & 0.850416 & 1.10063 & 1.23972 & 1.20093 & 1.50937 \\
$M/M_{\odot}$ eqn. (5) & 1.85688 & 1.24474 & 0.939454 & 1.43813 & 2.24996 & 1.85285 & 0.912947 & 0.943931 & 0.833866 & 1.00898 & 1.31735 \\
$M/M_{\odot}$ eqn. (6) & 1.56547 & 1.45892 & 1.01461 & 1.40432 & 1.78445 & 1.52981 & 0.891609 & 0.993515 & 0.951712 & 1.06929 & 1.37847 \\
$M/M_{\odot}$ eqn. (7) & 1.23267 & 1.82207 & 1.13003 & 1.35833 & 1.28993 & 1.16992 & 0.862570 & 1.06734 & 1.14519 & 1.15982 & 1.46885 \\
$M/M_{\odot}$ mean & 1.44192 & 1.63248 & 1.06688 & 1.38497 & 1.61170 & 1.39887 & 0.879386 & 1.02635 & 1.04262 & 1.10975 & 1.41851 \\
$\sigma_{M/M_{\odot}}$ & 0.336 & 0.343 & 0.110 & 0.0447 & 0.509 & 0.366 & 0.0282 & 0.0708 & 0.183 & 0.0868 & 0.0868 \\
\hline
\multicolumn{12}{l}{Assuming $(m-M)_V=13.51, E(B-V)=0.16$ and $G,(G_{\rm BP}-G_{\rm RP})$ photometry:}\\
$M/M_{\odot}$ eqn. (4) & 1.13287 & 1.90716 & 1.22998 & 1.30495 & 1.10672 & 1.04060 & 0.835151 & 1.08363 & 1.17851 & 1.17466 & 1.46863 \\
$M/M_{\odot}$ eqn. (5) & 1.82883 & 1.60053 & 0.826067 & 1.62553 & 2.39214 & 1.86040 & 0.980790 & 1.00440 & 1.13340 & 1.13266 & 1.49347 \\
$M/M_{\odot}$ eqn. (6) & 1.55899 & 1.69683 & 0.943286 & 1.51076 & 1.85013 & 1.53284 & 0.929620 & 1.03014 & 1.14824 & 1.14649 & 1.48514 \\
$M/M_{\odot}$ eqn. (7) & 1.24675 & 1.84146 & 1.13585 & 1.36356 & 1.29118 & 1.16882 & 0.862436 & 1.06730 & 1.16934 & 1.16613 & 1.47356 \\
$M/M_{\odot}$ mean & 1.44186 & 1.76150 & 1.03380 & 1.45120 & 1.66004 & 1.40066 & 0.901999 & 1.04637 & 1.15737 & 1.15498 & 1.48020 \\
$\sigma_{M/M_{\odot}}$ & 0.314 & 0.138 & 0.182 & 0.144 & 0.581 & 0.370 & 0.0658 & 0.0358 & 0.0203 & 0.0189 & 0.0112 \\
\hline
Our classification$^{13}$ & SNM & SOM & SM & SOM & SNM & BNM & SUM & SM & SM & SM & SOM \\
\hline
\noalign{\smallskip}
\multicolumn{12}{l}{$^1$ \citet{Stetson2003}. $^2$ \citet{Platais2011}. $^3$ \citet{Brogaard2012}. $^4$ \citet{Tofflemire2014}. }\\

\multicolumn{12}{l}{$^5$ Hyphens are not part of the ID, but only to allow ID to extend over three lines. $^6V$-mag from \citet{Stetson2003}, KIC\,2438139 from K. Cudworth (Priv. Comm.).}\\

\multicolumn{12}{l}{$^7$ APOGEE DR16 \citep{Majewski2017}. $^8$ \citet{Villanova2018}. 
$^9$ Gaia EDR3 parallax with \citet{Lindegren2021} zp correction. $^{10}$ $(m-M]_0$ from zp corr. par.
}\\

\multicolumn{12}{l}{$^{11}$ Evolutionary state from $\Delta P$, or alternative methods (marked). $^{12}$ From \citet{Rodrigues2017}.
}\\

\multicolumn{12}{l}{
$^{13}$ S: single, M: member, NM: non-member, O: overmassive, U: undermassive
}\\
\end{tabular}
\label{tab:starpars}
\end{table*}

\section{Results \& Discussion}
\label{sec:discussion}

The general results that emerge from interpretation of Table \ref{tab:starpars}, Fig.~\ref{fig:CMD}, and Fig.~\ref{fig:masspar} are the following: 

Three stars, KIC\,2570652, KIC\,2438139, and KIC\,2707478, are clear non-members located at a roughly common distance which is smaller than that of the cluster. This is evidenced by their Gaia EDR3 parallaxes, which are supported by the much smaller mass scatter among the results from the different mass equations when using the parallax than when assuming the cluster distance. Both radial velocity and proper motion also clearly imply non-membership.
All three non-members are low mass giants and thus old stars. Differences in radial velocities and proper motions indicate that they are most likely not related in any other way than their common distance. But the very similar properties of KIC\,2438139 to the cluster RC stars and the indication from Fig.~\ref{fig:CMD} that KIC\,270478 is an early AGB star with a mass consistent with the cluster, leaves a suspicion that these stars could be former cluster members. Comparsions to the isochrones in the upper panel of Fig.~\ref{fig:CMD} show that all stars can be assumed to have very close to the same reddening, even the non-members, except for KIC\,2570652, which cannot be naturally explained by an isochrone unless its reddening is significantly lower than that of the cluster, the metallicity is much lower than that of the cluster, and/or something has affected the photometry.

We consulted the 3D reddening map of \citet{Green2019} at the coordinates of our targets. With the exception of two stars they all yielded $E(g-r)$ between 0.14 and 0.17 with uncertainties of 0.02 at their parallax distance. KIC\,2570652, however, yields $E(g-r)=0.10$, about 0.055 lower. Using the transformation $E(B_V)=0.884\times E(g-r)$ from \citet{Green2019} and $E(V-K_S)=2.72\times E(B-V)$ from \citet{Casagrande2014} this would lower the reddening in $(V-K_S)$ by about 0.13. As seen in the top panel of Fig. 4, this makes the colour more consistent with stars of similar mass, when accounting for the fact that this is an RGB star with a radius similar to the cluster RGB members. However, the colour is still smaller than expected for the cluster, which is even more pronounced for the Gaia colour in the lower panel. Since $G_{\rm BP}-G_{\rm RP}$ is more sensitive to metallicity than $(V-K_S)$, this suggests that the metallicity of KIC\,2570652 is lower than NGC\,6791. 
Returning to the 3D reddening map of \citet{Green2019}, one of the other non-members, KIC\,2438139 shows a very discrepant and oddly low value of $E(g-r)=0.01$ at the parallax distance. At the same time, there is a steep increase to $E(g-r)=0.17$ at about the cluster distance. If the very low reddening was true then it would not be possible to match the colour of the star with any isochrone when accounting for the RC nature of the star unless it has a much higher metallicity then any other star known. We therefore suspect that the steep incline in reddening with distance in the 3D reddening map is due to a lack of stars a lower distances, and that the true reddening of KIC\,2438139 is similar to the reddening at the cluster distance.
The third non-member, KIC\,2707478 has a reddening map value of $E(g-r)=0.014$ largely independent of the assumed distance and consistent with that of the cluster. 

The potentially incorrect assumption of super-solar metallicity for the non-members has minor influence on their derived parameters and distances. Reducing the assumed metallicity from ${\rm [Fe/H]=+0.35}$ to ${\rm [Fe/H]=0.0}$ changes the bolometric corrections ${\rm BC}_G$ and ${\rm BC}_V$ by 0.005-0.001 mag and reduces the predicted colours by 0.022-0.046 mag for ($G_{\rm BP}-G_{\rm RP}$) and 0.013-0.028 mag for $(V-K_S)$. These colour changes correspond to $T_{\rm eff}$ changes of 37-77 K for the Gaia colour and 7-19 K for $(V-K_S)$. The isochrone comparison thus supports that KIC\,2438139 and KIC\,2707478 have properties consistent with the cluster members, except for the distance, suggesting that perhaps they are past cluster members. High resolution follow-up spectroscopy is needed to investigate this possibility through abundance measurements.

The other giants are all cluster members based on their parallaxes, radial velocities, and proper motions.
However, while some members have similar masses close to what is expected for giant stars in the cluster, $M \sim 1.15 M_{\odot}$ on the RGB \citep{Brogaard2012} and about $0.09 M_{\odot}$ less in the RC \citep{Miglio2012}, others are significantly more massive, as seen in Fig.~\ref{fig:masspar}. This indicates that they experienced mass-transfer and/or a stellar merger in their past.

As part of our analysis, we calculated the true distance modulus from the Gaia EDR3 parallaxes with corrected zeropoint. The weighted mean for the cluster members is $\pi=0.237\pm0.006$ mmag corresponding to $(m-M)_0=13.13\pm0.05$ mag with an additional systematic $\pm0.13$ mag due to the potential additional 0.015 mmag Gaia EDR3 parallax zeropoint offset determined by \citet{Stassun2021}. This can be considered an upper limit to the systematic uncertainty, since other studies find parallax offsets of this size or smaller \citep{Huang2021,Ren2021,Vasiliev2021,Zinn2021} after the correction by \citet{Lindegren2021}, but not in the same direction.

In Fig.~\ref{fig:masspar} we compare the masses and parallaxes to predictions of cluster RGB mass and parallax from eclipsing binary members of NGC\,6791 \citep{Brogaard2012}. For the parallax prediction, we have adopted $E(B-V)=0.16$ for the estimate, but allowed for the possibility of $E(B-V)=0.10$ in the uncertainty estimate. We also compare to a predicted parallax from our asteroseismic measurements and photometric $T_{\rm eff}$ values. 
Taken at face values, the comparison to the eclipsing binary predictions suggest that NGC\,6791 has $E(B-V)\simeq0.13$. However, we cannot discriminate strongly between this and our adopted $E(B-V)=0.16$ or even the lower extreme $E(B-V)=0.10$ due to the combined effects of uncertainties in the parallax, the parallax zeropoint, the $T_{\rm eff}$ scale, and the adopted independent apparent distance modulus. Our asteroseismic parallax estimates also shown in Fig.~\ref{fig:masspar} seem to support the Gaia parallaxes, and thus $E(B-V)\simeq0.13$, but they depend critically on the adopted bolometric corrections, and also on $E(B-V)$ with lower values resulting in larger parallax predictions. We therefore refrain from trying to optimize $E(B-V)$ though this might be possible in the future if reddening independent spectroscopic $T_{\rm eff}$ values and precise asteroseismic parameters can be obtained for a larger sample of cluster giants. Luckily, our conclusions are robust against the adopted $E(B-V)$, except for small changes to the absolute mass scale.  

\section{Details on individual targets}

In this section we give details on the individual targets based on the derived properties from Tables~\ref{tab:seis} and~\ref{tab:starpars}, and Fig.~\ref{fig:CMD}.

\subsection{Undermassive cluster member}

KIC\,2436543 was originally chosen to serve as a reference of the normal cluster stars that evolved as single stars. As it turned out, the star is a clear member according to RV, PM, and parallax, but the mass, $M=0.90\pm0.07 M_{\odot}$, is lower than measured for any other RC star in the cluster - both in our study and all others in the literature, signaling that for some reason this star experienced a higher mass-loss than any other measured RC giant in this cluster. In this sense it resembles the very low mass $0.71\pm0.08 M_{\odot}$ RC star KIC\,4937011 in NGC\,6819 measured and discussed by \citet{Handberg2017}, although the mass loss is less extreme in our case. KIC\,4937011 is also Li-rich \citep{Anthony-Twarog2013}. Therefore, whatever caused the high mass-loss might also have caused Li production. Investigating whether KIC\,2436543 is Li-rich might therefore reveal further resemblance, or not, to KIC\,4937011.

\subsection{Overmassive cluster members}

KIC\,2437267 is a clear member according to RV, PM, and parallax. It is an overmassive $M=1.45\pm0.15 M_{\odot}$ RC star according to asteroseismology.

KIC\,2437589 was not in the study by \citet{Platais2011} but noted as an outlier in \citet{Corsaro2012} and suggested as an overmassive RGB member by \citet{Brogaard2012}.
We could not measure an asteroseismic $\Delta P$ and it is therefore not certain whether it is an RGB or RC star. We choose the RGB phase, since this gives the smallest mass scatter, $M=1.49\pm0.02 M_{\odot}$, and it is also the only phase consistent with the CMD position for a star of this mass in Fig.~\ref{fig:CMD}.

KIC\,2438100, is most likely a cluster member, but the interpretation is a little unclear. 
The radial velocity membership probability is relatively low (56\%) and the proper motion values are at the limits of what can be considered for members, but also not ruling out membership; Gaia DR2 (PMRA,PMDEC) = (-0.508,-2.193) compared to the mean cluster values of (-0.421,-2.269) \citep{Cantat2018} or ($-0.434\pm0.008,-2.266\pm0.010$) \citet{Gao2020}. 
The Gaia EDR3 parallax is consistent with cluster membership, but the parallax changed significantly between DR2 and EDR3, suggesting some complication.
We could not measure an asteroseismic period spacing $\Delta P$ to determine the evolutionary status. Without this classification there are two possibilities for the correction to $\Delta \nu$, yielding either an RGB star with a mass of $M=1.73\pm0.06 M_{\odot}$, or a clump star with a mass of $1.83\pm0.14 M_{\odot}$.   

The assumption of the RGB phase results in a lower mass scatter by a factor of about two compared to a RC scenario. However, the CMD position suggests a helium-burning star and so does $\epsilon_{\rm c}$, as explained earlier, which is why we favour the RC scenario. The derived parallax is then 0.212 mas, which is the lowest of the members. However, the effective temperatures from the two colours are different by more than 160 K, signalling large uncertainty, and if $T_{\rm eff}$ is actually a bit lower than the smallest estimate then the derived parallax is identical to the Gaia EDR3 parallax. 

A metallicity measurement would be of great value for fully decisive conclusion. We were unable to find a high-resolution result in the literature, but \citet{Warren2009} has a calcium triplet (CaT) equivalent width measurements of this star and other NGC6791 giants. This is however not conclusive, since KIC\,2438100 has the lowest value among their NGC\,6791 targets. While this is partly expected due to a known correlation between $K_s$ magnitude and the CaT, we note that their CaT measurement values for the solar metallicity cluster NGC\,6819 \citep{Slumstrup2019} are as high as those for NGC\,6791, making this a weak membership discriminator.

Since all membership criteria RV, PM, parallax, and CaT suggest borderline membership, we assume this star is a member.
The mass is fairly high and one could speculate that some kind of violent mass transfer or merger has been involved. This could perhaps naturally explain why the RV, PM, and parallax are close to but not exactly as expected for a member, since a collision could have affected these parameters.

\subsection{Normal cluster members}

Asteroseismology places KIC\,2436944 in the RC phase. The RV suggests membership, and the mass is $M=1.03\pm0.18 M_{\odot}$, as expected for a normal RC member. The Gaia DR2 showed a negative parallax and no PM measurements. Although this was corrected in EDR3, this target has a nearby companion, which likely still affects the Gaia parameters, including the photometry. Supporting this is the fact that in Fig.~\ref{fig:CMD}, the $(V-K_s)$ colour seems much more consistent with the RC location than the Gaia colours. Also, the mass scatter is much smaller when assuming the eclipsing binary based cluster distance rather than the Gaia EDR3 parallax.
Since nothing was found out of the ordinary, except for the uncertain Gaia parallax that is likely caused by a nearby star, we suggest that the very same issue of erroneous photometry is what caused \citet{Platais2011} to select this star as being special.

KIC\,2437353 has properties very similar to KIC\,2436944 in all aspects, including the mass, which is $M=1.05\pm0.05 M_{\odot}$, which is expected for a RC member that evolved as single.

KIC\,2570094 and KIC\,2438140 are two very similar RGB members as evident from both the CMDs and their asteroseseimic parameters. Their masses can be assumed to be identical at the measurement precision level, and are $M=1.20\pm0.01 M_{\odot}$ if adopting the Gaia parallax, $M=1.08\pm0.04 M_{\odot}$ using the $V,(V-K_s)$ photometry and binary distance, or $M=1.16\pm0.01 M_{\odot}$ using the $G,(G_{\rm BP}-G_{\rm RP})$ photometry and binary distance. 

\subsection{Non-members}


KIC\,2570652 is on the lower RGB with a mass just below 1 $M_{\odot}$. Since no measurable mass-loss is expected at the early giant phase the low mass indicates an old star.

KIC\,2438139 is classified as RC and has very similar parameters to the RC members of the cluster. The mass of $M=1.12\pm0.01 M_{\odot}$ as derived using the Gaia EDR3 parallax is also consistent with that of the RC members. However, the parallax is significantly larger than for the cluster members and the magnitudes are smaller than the member RC stars in the CMDs, both signalling that the star is closer than the cluster. RVs and PMs clearly indicate non-membership. High resolution spectroscopic follow-up is needed to investigate element abundances that can reveal whether this star was once a part of NGC\,6791. 

KIC\,2707478 is similarly just around 1 $M_{\odot}$, and not member according to RV, RM and parallax. Global seismic parameters are consistent with low-mass red clump, but we have no $\Delta P$ measurement to confirm that. The RC phase was chosen because of a smaller mass scatter than if the RGB phase was adopted. The radius then suggests that the star is in the late RC or early AGB phase, which is also consistent with the CMD position in Fig~\ref{fig:CMD}.
SB1 Binarity was detected by \citet{Tofflemire2014}. This could have affected the photometric measurements.

\section{non-standard stellar evolution products}
\label{sec:non-standard}

The main purpose of our investigation was to look for the existence of overmassive cluster members in NGC\,6791 among the photometric outliers identified by \citet{Platais2011}. Three such stars are now identified, KIC\,2438100, KIC\,2437267, and KIC\,2437589. It is difficult to turn this into a percentage of cluster stars that turn overmassive during their evolution. One way to obtain a percentage is to take the number of overmassive giants relative to the number of RC stars, since only this area of the CMD has been searched for overmassive stars.
A problem that arises is that we can only do asteroseismic measurements for relatively bright giants in the cluster, and that there are in fact still quite a few of them that have not (yet) been investigated. In Fig.~\ref{fig:CMD}, the RC stars that have been investigated asteroseismically by \citet{Bossini2020} are marked with black open circles, but it is clear that more RC stars are present in the cluster. Searching the literature did not reveal additional RC stars with asterosismic measurements.  
A number count in the observed CMD by \citet{Cantat2018} gives about 30 stars consistent with the RC phase, while \citet{Gao2020} lists 27 RC stars. Using this, the number of cluster stars that turn overmassive would be roughly 3/30 = 10 percent. This is however is a very uncertain number because there might be more overmassive stars among those that have not been investigated yet. An additional complication is caused by the EHB stars present in NGC\,6791 \citep{Liebert1994}, which are also thought to be related to mass-transfer in binary stars. There are about 6 such stars in the cluster, as we verified using TOPCAT \citep{Taylor2005} to select proper motion members in NGC6791. The same number has been confirmed through spectroscopy \citep{Liebert1994}, three also thorough asteroseismology \citep{Reed2012}. If we include those in our calculation, we end up with only 3/36 = 8,3 percent of stars turning overmassive. There is also the complication of the mass dependence of the lifetimes of giants discussed by \citet{Miglio2021}, which we have made no attempt to account for.

Despite these complications, a similar investigation in the open cluster NGC\,6819 yielded a similar fraction, 6/51=12\%, of overmassive giants \citep{Handberg2017}, suggesting that a number close to 10 percent might be common among open clusters, and therefore perhaps also common among field stars. This would be consistent with the 9-14\% overmassive field giants identified in the APOKASC \citep{Pinsonneault2018} sample of thick disc stars by \citet{Izzard2018}. 

However, the situation might be more complicated than suggested by the similar numbers from the two open clusters and the asteroseismic field giant sample. The three overmassive giants in NGC\,6791 all appear to be single stars \citep{Tofflemire2014}, whereas most of the overmassive giants in NGC\,6819 are in eccentric long period binary systems with periods of the order $P\sim1000$ days \citep{Handberg2017,Milliman2014}, indicating that they were perhaps triples at some point. Therefore, it is not clear whether the formation mechanisms and evolutionary scenarios are similar.

\section{Conclusions}
\label{sec:conclusion}

We performed an asteroseismic investigation of giant stars in the field of NGC\,6791 with previous indications of atypical evolution. Among these stars we found evidence of overmassive and undermassive cluster stars, and non-members with hints of potential past membership.

Our results show that about 10\% of red giants are expected to have experienced mass-transfer or a merger if the field population of the {Galaxy} is similar to that of the cluster. This is important, since for such stars the common age dating methods that assume single star evolution are invalid.

High-resolution high S/N spectroscopic follow-up could reveal the potential past membership of the non-members and determine whether some element abundances might allow to expose the non-standard evolution of overmassive and undermassive stars. If so, field stars of similar type could be identified as what they are, and not be mistakenly classified as younger or older than they are.

\section*{Acknowledgements}

We gratefully acknowledge the grant from the European Social Fund via the Lithuanian Science Council (LMTLT) grant No. 09.3.3-LMT-K-712-01-0103.\\
Funding for the Stellar Astrophysics Centre is provided by The Danish National Research Foundation (Grant agreement no.: DNRF106).\\
AM acknowledges funding from the European Research Council (ERC) under the European Union’s Horizon 2020 research and innovation programme (grant agreement No. 772293  - project ASTEROCHRONOMETRY, \url{https://www.asterochronometry.eu})\\
This research has made use of the SIMBAD database,
operated at CDS, Strasbourg, France\\
This research made use of Lightkurve, a Python package for Kepler and TESS data analysis (Lightkurve Collaboration, 2018).\\
This work has made use of data from the European Space Agency (ESA) mission
{\it Gaia} (\url{https://www.cosmos.esa.int/gaia}), processed by the {\it Gaia}
Data Processing and Analysis Consortium (DPAC,
\url{https://www.cosmos.esa.int/web/gaia/dpac/consortium}). Funding for the DPAC
has been provided by national institutions, in particular the institutions
participating in the {\it Gaia} Multilateral Agreement.\\

\section*{Data availability}

All data underlying this article are either available in the article or through the references cited. The only exception is the measured oscillation frequencies, which will be made available upon request.




\bibliographystyle{mnras}
\bibliography{NGC6791-refs.bib} 

\begin{thebibliography}{}
\makeatletter
\relax
\def\mn@urlcharsother{\let\do\@makeother \do\$\do\&\do\#\do\^\do\_\do\%\do\~}
\def\mn@doi{\begingroup\mn@urlcharsother \@ifnextchar [ {\mn@doi@}
  {\mn@doi@[]}}
\def\mn@doi@[#1]#2{\def\@tempa{#1}\ifx\@tempa\@empty \href
  {http://dx.doi.org/#2} {doi:#2}\else \href {http://dx.doi.org/#2} {#1}\fi
  \endgroup}
\def\mn@eprint#1#2{\mn@eprint@#1:#2::\@nil}
\def\mn@eprint@arXiv#1{\href {http://arxiv.org/abs/#1} {{\tt arXiv:#1}}}
\def\mn@eprint@dblp#1{\href {http://dblp.uni-trier.de/rec/bibtex/#1.xml}
  {dblp:#1}}
\def\mn@eprint@#1:#2:#3:#4\@nil{\def\@tempa {#1}\def\@tempb {#2}\def\@tempc
  {#3}\ifx \@tempc \@empty \let \@tempc \@tempb \let \@tempb \@tempa \fi \ifx
  \@tempb \@empty \def\@tempb {arXiv}\fi \@ifundefined
  {mn@eprint@\@tempb}{\@tempb:\@tempc}{\expandafter \expandafter \csname
  mn@eprint@\@tempb\endcsname \expandafter{\@tempc}}}

\bibitem[\protect\citeauthoryear{{An}, {Terndrup}, {Pinsonneault}  \&
  {Lee}}{{An} et~al.}{2015}]{An2015}
{An} D.,  {Terndrup} D.~M.,  {Pinsonneault} M.~H.,   {Lee} J.-W.,  2015,
  \mn@doi [\apj] {10.1088/0004-637X/811/1/46}, \href
  {https://ui.adsabs.harvard.edu/abs/2015ApJ...811...46A} {811, 46}

\bibitem[\protect\citeauthoryear{{Anthony-Twarog}, {Deliyannis}, {Rich}  \&
  {Twarog}}{{Anthony-Twarog} et~al.}{2013}]{Anthony-Twarog2013}
{Anthony-Twarog} B.~J.,  {Deliyannis} C.~P.,  {Rich} E.,   {Twarog} B.~A.,
  2013, \mn@doi [\apjl] {10.1088/2041-8205/767/1/L19}, \href
  {https://ui.adsabs.harvard.edu/abs/2013ApJ...767L..19A} {767, L19}

\bibitem[\protect\citeauthoryear{{Arentoft}, {Brogaard}, {Jessen-Hansen},
  {Silva Aguirre}, {Kjeldsen}, {Mosumgaard}  \& {Sandquist}}{{Arentoft}
  et~al.}{2017}]{Arentoft17}
{Arentoft} T.,  {Brogaard} K.,  {Jessen-Hansen} J.,  {Silva Aguirre} V.,
  {Kjeldsen} H.,  {Mosumgaard} J.~R.,   {Sandquist} E.~L.,  2017, \mn@doi
  [\apj] {10.3847/1538-4357/aa63f7}, \href
  {https://ui.adsabs.harvard.edu/abs/2017ApJ...838..115A} {838, 115}

\bibitem[\protect\citeauthoryear{{Arentoft} et~al.,}{{Arentoft}
  et~al.}{2019}]{Arentoft19}
{Arentoft} T.,  et~al., 2019, \mn@doi [\aap] {10.1051/0004-6361/201834690},
  \href {https://ui.adsabs.harvard.edu/abs/2019A&A...622A.190A} {622, A190}

\bibitem[\protect\citeauthoryear{{Bailer-Jones}, {Rybizki}, {Fouesneau},
  {Mantelet}  \& {Andrae}}{{Bailer-Jones} et~al.}{2018}]{Bailer2018}
{Bailer-Jones} C.~A.~L.,  {Rybizki} J.,  {Fouesneau} M.,  {Mantelet} G.,
  {Andrae} R.,  2018, \mn@doi [\aj] {10.3847/1538-3881/aacb21}, \href
  {https://ui.adsabs.harvard.edu/abs/2018AJ....156...58B} {156, 58}

\bibitem[\protect\citeauthoryear{{Basu} et~al.,}{{Basu}
  et~al.}{2011}]{Basu2011}
{Basu} S.,  et~al., 2011, \mn@doi [\apjl] {10.1088/2041-8205/729/1/L10}, \href
  {https://ui.adsabs.harvard.edu/abs/2011ApJ...729L..10B} {729, L10}

\bibitem[\protect\citeauthoryear{{Bedding} et~al.,}{{Bedding}
  et~al.}{2011}]{Bedding11}
{Bedding} T.~R.,  et~al., 2011, \mn@doi [\nat] {10.1038/nature09935}, \href
  {https://ui.adsabs.harvard.edu/abs/2011Natur.471..608B} {471, 608}

\bibitem[\protect\citeauthoryear{{Belkacem}, {Goupil}, {Dupret}, {Samadi},
  {Baudin}, {Noels}  \& {Mosser}}{{Belkacem} et~al.}{2011}]{Belkacem2011}
{Belkacem} K.,  {Goupil} M.~J.,  {Dupret} M.~A.,  {Samadi} R.,  {Baudin} F.,
  {Noels} A.,   {Mosser} B.,  2011, \mn@doi [\aap]
  {10.1051/0004-6361/201116490}, \href
  {https://ui.adsabs.harvard.edu/abs/2011A&A...530A.142B} {530, A142}

\bibitem[\protect\citeauthoryear{{Borucki} et~al.,}{{Borucki}
  et~al.}{2010}]{Borucki2010}
{Borucki} W.~J.,  et~al., 2010, \mn@doi [Science] {10.1126/science.1185402},
  \href {https://ui.adsabs.harvard.edu/abs/2010Sci...327..977B} {327, 977}

\bibitem[\protect\citeauthoryear{{Bossini} et~al.,}{{Bossini}
  et~al.}{2020}]{Bossini2020}
{Bossini} D.,  et~al., 2020, VizieR Online Data Catalog, \href
  {https://ui.adsabs.harvard.edu/abs/2020yCat..74694718B} {p. J/MNRAS/469/4718}

\bibitem[\protect\citeauthoryear{{Bressan}, {Marigo}, {Girardi}, {Salasnich},
  {Dal Cero}, {Rubele}  \& {Nanni}}{{Bressan} et~al.}{2012}]{Bressan2012}
{Bressan} A.,  {Marigo} P.,  {Girardi} L.,  {Salasnich} B.,  {Dal Cero} C.,
  {Rubele} S.,   {Nanni} A.,  2012, \mn@doi [\mnras]
  {10.1111/j.1365-2966.2012.21948.x}, \href
  {https://ui.adsabs.harvard.edu/abs/2012MNRAS.427..127B} {427, 127}

\bibitem[\protect\citeauthoryear{{Brogaard}, {Bruntt}, {Grundahl}, {Clausen},
  {Frandsen}, {Vandenberg}  \& {Bedin}}{{Brogaard} et~al.}{2011}]{Brogaard2011}
{Brogaard} K.,  {Bruntt} H.,  {Grundahl} F.,  {Clausen} J.~V.,  {Frandsen} S.,
  {Vandenberg} D.~A.,   {Bedin} L.~R.,  2011, \mn@doi [\aap]
  {10.1051/0004-6361/201015503}, \href
  {https://ui.adsabs.harvard.edu/abs/2011A&A...525A...2B} {525, A2}

\bibitem[\protect\citeauthoryear{{Brogaard} et~al.,}{{Brogaard}
  et~al.}{2012}]{Brogaard2012}
{Brogaard} K.,  et~al., 2012, \mn@doi [\aap] {10.1051/0004-6361/201219196},
  \href {https://ui.adsabs.harvard.edu/abs/2012A&A...543A.106B} {543, A106}

\bibitem[\protect\citeauthoryear{{Brogaard} et~al.,}{{Brogaard}
  et~al.}{2016}]{Brogaard2016}
{Brogaard} K.,  et~al., 2016, \mn@doi [Astronomische Nachrichten]
  {10.1002/asna.201612374}, \href
  {https://ui.adsabs.harvard.edu/abs/2016AN....337..793B} {337, 793}

\bibitem[\protect\citeauthoryear{{Brogaard} et~al.,}{{Brogaard}
  et~al.}{2018a}]{Brogaard2018dEBs}
{Brogaard} K.,  et~al., 2018a, \mn@doi [\mnras] {10.1093/mnras/sty268}, \href
  {https://ui.adsabs.harvard.edu/abs/2018MNRAS.476.3729B} {476, 3729}

\bibitem[\protect\citeauthoryear{{Brogaard} et~al.,}{{Brogaard}
  et~al.}{2018b}]{Brogaard2018}
{Brogaard} K.,  et~al., 2018b, \mn@doi [\mnras] {10.1093/mnras/sty2504}, \href
  {https://ui.adsabs.harvard.edu/abs/2018MNRAS.481.5062B} {481, 5062}

\bibitem[\protect\citeauthoryear{{Cantat-Gaudin} et~al.,}{{Cantat-Gaudin}
  et~al.}{2018}]{Cantat2018}
{Cantat-Gaudin} T.,  et~al., 2018, \mn@doi [\aap]
  {10.1051/0004-6361/201833476}, \href
  {https://ui.adsabs.harvard.edu/abs/2018A&A...618A..93C} {618, A93}

\bibitem[\protect\citeauthoryear{{Casagrande} \& {VandenBerg}}{{Casagrande} \&
  {VandenBerg}}{2014}]{Casagrande2014}
{Casagrande} L.,  {VandenBerg} D.~A.,  2014, \mn@doi [\mnras]
  {10.1093/mnras/stu1476}, \href
  {https://ui.adsabs.harvard.edu/abs/2014MNRAS.444..392C} {444, 392}

\bibitem[\protect\citeauthoryear{{Casagrande} \& {VandenBerg}}{{Casagrande} \&
  {VandenBerg}}{2018}]{Casagrande2018}
{Casagrande} L.,  {VandenBerg} D.~A.,  2018, \mn@doi [\mnras]
  {10.1093/mnrasl/sly104}, \href
  {https://ui.adsabs.harvard.edu/abs/2018MNRAS.479L.102C} {479, L102}

\bibitem[\protect\citeauthoryear{{Chiappini} et~al.,}{{Chiappini}
  et~al.}{2015}]{Chiappini2015}
{Chiappini} C.,  et~al., 2015, \mn@doi [\aap] {10.1051/0004-6361/201525865},
  \href {https://ui.adsabs.harvard.edu/abs/2015A&A...576L..12C} {576, L12}

\bibitem[\protect\citeauthoryear{{Christensen-Dalsgaard}, {Arentoft}, {Brown},
  {Gilliland}, {Kjeldsen}, {Borucki}  \& {Koch}}{{Christensen-Dalsgaard}
  et~al.}{2008}]{JCD08}
{Christensen-Dalsgaard} J.,  {Arentoft} T.,  {Brown} T.~M.,  {Gilliland} R.~L.,
   {Kjeldsen} H.,  {Borucki} W.~J.,   {Koch} D.,  2008, in Journal of Physics
  Conference Series. p. 012039, \mn@doi{10.1088/1742-6596/118/1/012039}

\bibitem[\protect\citeauthoryear{{Corsaro} et~al.,}{{Corsaro}
  et~al.}{2012}]{Corsaro2012}
{Corsaro} E.,  et~al., 2012, \mn@doi [\apj] {10.1088/0004-637X/757/2/190},
  \href {https://ui.adsabs.harvard.edu/abs/2012ApJ...757..190C} {757, 190}

\bibitem[\protect\citeauthoryear{{Gaia Collaboration}, {Brown}, {Vallenari},
  {Prusti}, {de Bruijne}, {Babusiaux}  \& {Biermann}}{{Gaia Collaboration}
  et~al.}{2020}]{Gaiamission2020}
{Gaia Collaboration} {Brown} A.~G.~A.,  {Vallenari} A.,  {Prusti} T.,  {de
  Bruijne} J.~H.~J.,  {Babusiaux} C.,   {Biermann} M.,  2020, arXiv e-prints,
  \href {https://ui.adsabs.harvard.edu/abs/2020arXiv201201533G} {p.
  arXiv:2012.01533}

\bibitem[\protect\citeauthoryear{{Gao}}{{Gao}}{2020}]{Gao2020}
{Gao} X.,  2020, \mn@doi [\apss] {10.1007/s10509-020-3738-2}, \href
  {https://ui.adsabs.harvard.edu/abs/2020Ap&SS.365...24G} {365, 24}

\bibitem[\protect\citeauthoryear{{Green}, {Schlafly}, {Zucker}, {Speagle}  \&
  {Finkbeiner}}{{Green} et~al.}{2019}]{Green2019}
{Green} G.~M.,  {Schlafly} E.,  {Zucker} C.,  {Speagle} J.~S.,   {Finkbeiner}
  D.,  2019, \mn@doi [\apj] {10.3847/1538-4357/ab5362}, \href
  {https://ui.adsabs.harvard.edu/abs/2019ApJ...887...93G} {887, 93}

\bibitem[\protect\citeauthoryear{{Handberg} \& {Lund}}{{Handberg} \&
  {Lund}}{2014}]{Handberg2014}
{Handberg} R.,  {Lund} M.~N.,  2014, \mn@doi [\mnras] {10.1093/mnras/stu1823},
  \href {https://ui.adsabs.harvard.edu/abs/2014MNRAS.445.2698H} {445, 2698}

\bibitem[\protect\citeauthoryear{{Handberg}, {Brogaard}, {Miglio}, {Bossini},
  {Elsworth}, {Slumstrup}, {Davies}  \& {Chaplin}}{{Handberg}
  et~al.}{2017}]{Handberg2017}
{Handberg} R.,  {Brogaard} K.,  {Miglio} A.,  {Bossini} D.,  {Elsworth} Y.,
  {Slumstrup} D.,  {Davies} G.~R.,   {Chaplin} W.~J.,  2017, \mn@doi [\mnras]
  {10.1093/mnras/stx1929}, \href
  {https://ui.adsabs.harvard.edu/abs/2017MNRAS.472..979H} {472, 979}

\bibitem[\protect\citeauthoryear{{Holtzman} et~al.,}{{Holtzman}
  et~al.}{2018}]{Holtzman2018}
{Holtzman} J.~A.,  et~al., 2018, \mn@doi [\aj] {10.3847/1538-3881/aad4f9},
  \href {https://ui.adsabs.harvard.edu/abs/2018AJ....156..125H} {156, 125}

\bibitem[\protect\citeauthoryear{{Huang}, {Yuan}, {Beers}  \& {Zhang}}{{Huang}
  et~al.}{2021}]{Huang2021}
{Huang} Y.,  {Yuan} H.,  {Beers} T.~C.,   {Zhang} H.,  2021, \mn@doi [\apjl]
  {10.3847/2041-8213/abe69a}, \href
  {https://ui.adsabs.harvard.edu/abs/2021ApJ...910L...5H} {910, L5}

\bibitem[\protect\citeauthoryear{{Izzard}, {Preece}, {Jofre}, {Halabi},
  {Masseron}  \& {Tout}}{{Izzard} et~al.}{2018}]{Izzard2018}
{Izzard} R.~G.,  {Preece} H.,  {Jofre} P.,  {Halabi} G.~M.,  {Masseron} T.,
  {Tout} C.~A.,  2018, \mn@doi [\mnras] {10.1093/mnras/stx2355}, \href
  {https://ui.adsabs.harvard.edu/abs/2018MNRAS.473.2984I} {473, 2984}

\bibitem[\protect\citeauthoryear{{Jofr{\'e}} et~al.,}{{Jofr{\'e}}
  et~al.}{2016}]{Jofre2016}
{Jofr{\'e}} P.,  et~al., 2016, \mn@doi [\aap] {10.1051/0004-6361/201629356},
  \href {https://ui.adsabs.harvard.edu/abs/2016A&A...595A..60J} {595, A60}

\bibitem[\protect\citeauthoryear{{J{\"o}nsson} et~al.,}{{J{\"o}nsson}
  et~al.}{2020}]{Jonsson2020}
{J{\"o}nsson} H.,  et~al., 2020, \mn@doi [\aj] {10.3847/1538-3881/aba592},
  \href {https://ui.adsabs.harvard.edu/abs/2020AJ....160..120J} {160, 120}

\bibitem[\protect\citeauthoryear{{Kallinger} et~al.,}{{Kallinger}
  et~al.}{2012}]{Kallinger12}
{Kallinger} T.,  et~al., 2012, \mn@doi [\aap] {10.1051/0004-6361/201218854},
  \href {https://ui.adsabs.harvard.edu/abs/2012A&A...541A..51K} {541, A51}

\bibitem[\protect\citeauthoryear{{Kinman}}{{Kinman}}{1965}]{Kinman1965}
{Kinman} T.~D.,  1965, \mn@doi [\apj] {10.1086/148329}, \href
  {https://ui.adsabs.harvard.edu/abs/1965ApJ...142..655K} {142, 655}

\bibitem[\protect\citeauthoryear{{Kjeldsen} \& {Bedding}}{{Kjeldsen} \&
  {Bedding}}{1995}]{Kjeldsen1995}
{Kjeldsen} H.,  {Bedding} T.~R.,  1995, \aap, \href
  {https://ui.adsabs.harvard.edu/abs/1995A&A...293...87K} {293, 87}

\bibitem[\protect\citeauthoryear{{Kuehn}, {Drury}, {Bellamy}, {Stello},
  {Bedding}, {Reed}  \& {Quick}}{{Kuehn} et~al.}{2015}]{Kuehn2015}
{Kuehn} C.~A.,  {Drury} J.~A.,  {Bellamy} B.~R.,  {Stello} D.,  {Bedding}
  T.~R.,  {Reed} M.,   {Quick} B.,  2015, in European Physical Journal Web of
  Conferences. p. 06040, \mn@doi{10.1051/epjconf/201510106040}

\bibitem[\protect\citeauthoryear{{Liebert}, {Saffer}  \& {Green}}{{Liebert}
  et~al.}{1994}]{Liebert1994}
{Liebert} J.,  {Saffer} R.~A.,   {Green} E.~M.,  1994, \mn@doi [\aj]
  {10.1086/116954}, \href
  {https://ui.adsabs.harvard.edu/abs/1994AJ....107.1408L} {107, 1408}

\bibitem[\protect\citeauthoryear{{Lightkurve Collaboration}
  et~al.,}{{Lightkurve Collaboration} et~al.}{2018}]{LCcollab2018}
{Lightkurve Collaboration} et~al., 2018, {Lightkurve: Kepler and TESS time
  series analysis in Python}, Astrophysics Source Code Library (\mn@eprint
  {ascl} {1812.013})

\bibitem[\protect\citeauthoryear{{Lindegren} et~al.,}{{Lindegren}
  et~al.}{2021}]{Lindegren2021}
{Lindegren} L.,  et~al., 2021, \mn@doi [\aap] {10.1051/0004-6361/202039653},
  \href {https://ui.adsabs.harvard.edu/abs/2021A&A...649A...4L} {649, A4}

\bibitem[\protect\citeauthoryear{{Majewski} et~al.,}{{Majewski}
  et~al.}{2017}]{Majewski2017}
{Majewski} S.~R.,  et~al., 2017, \mn@doi [\aj] {10.3847/1538-3881/aa784d},
  \href {https://ui.adsabs.harvard.edu/abs/2017AJ....154...94M} {154, 94}

\bibitem[\protect\citeauthoryear{{Martig} et~al.,}{{Martig}
  et~al.}{2015}]{Martig2015}
{Martig} M.,  et~al., 2015, \mn@doi [\mnras] {10.1093/mnras/stv1071}, \href
  {https://ui.adsabs.harvard.edu/abs/2015MNRAS.451.2230M} {451, 2230}

\bibitem[\protect\citeauthoryear{{Miglio} et~al.,}{{Miglio}
  et~al.}{2012}]{Miglio2012}
{Miglio} A.,  et~al., 2012, \mn@doi [\mnras]
  {10.1111/j.1365-2966.2011.19859.x}, \href
  {https://ui.adsabs.harvard.edu/abs/2012MNRAS.419.2077M} {419, 2077}

\bibitem[\protect\citeauthoryear{{Miglio} et~al.,}{{Miglio}
  et~al.}{2021}]{Miglio2021}
{Miglio} A.,  et~al., 2021, \mn@doi [\aap] {10.1051/0004-6361/202038307}, \href
  {https://ui.adsabs.harvard.edu/abs/2021A&A...645A..85M} {645, A85}

\bibitem[\protect\citeauthoryear{{Milliman}, {Mathieu}, {Geller}, {Gosnell},
  {Meibom}  \& {Platais}}{{Milliman} et~al.}{2014}]{Milliman2014}
{Milliman} K.~E.,  {Mathieu} R.~D.,  {Geller} A.~M.,  {Gosnell} N.~M.,
  {Meibom} S.,   {Platais} I.,  2014, \mn@doi [\aj]
  {10.1088/0004-6256/148/2/38}, \href
  {https://ui.adsabs.harvard.edu/abs/2014AJ....148...38M} {148, 38}

\bibitem[\protect\citeauthoryear{{Montalb{\'a}n}, {Miglio}, {Noels},
  {Scuflaire}, {Ventura}  \& {D'Antona}}{{Montalb{\'a}n}
  et~al.}{2012}]{Montalban2012}
{Montalb{\'a}n} J.,  {Miglio} A.,  {Noels} A.,  {Scuflaire} R.,  {Ventura} P.,
   {D'Antona} F.,  2012, \mn@doi [Astrophysics and Space Science Proceedings]
  {10.1007/978-3-642-18418-5\_3}, \href
  {https://ui.adsabs.harvard.edu/abs/2012ASSP...26...23M} {26, 23}

\bibitem[\protect\citeauthoryear{{Pinsonneault} et~al.,}{{Pinsonneault}
  et~al.}{2018}]{Pinsonneault2018}
{Pinsonneault} M.~H.,  et~al., 2018, \mn@doi [\apjs]
  {10.3847/1538-4365/aaebfd}, \href
  {https://ui.adsabs.harvard.edu/abs/2018ApJS..239...32P} {239, 32}

\bibitem[\protect\citeauthoryear{{Platais}, {Cudworth}, {Platais-Kozhurina},
  {McLaughlin}, {Meibom}  \& {Veillet}}{{Platais} et~al.}{2011}]{Platais2011}
{Platais} I.,  {Cudworth} K.~M.,  {Platais-Kozhurina} V.,  {McLaughlin} D.~E.,
  {Meibom} S.,   {Veillet} C.,  2011, in American Astronomical Society Meeting
  Abstracts \#218. p. 133.02

\bibitem[\protect\citeauthoryear{{Reed}, {Baran}, {{\O}stensen}, {Telting}  \&
  {O'Toole}}{{Reed} et~al.}{2012}]{Reed2012}
{Reed} M.~D.,  {Baran} A.,  {{\O}stensen} R.~H.,  {Telting} J.,   {O'Toole}
  S.~J.,  2012, \mn@doi [\mnras] {10.1111/j.1365-2966.2012.22054.x}, \href
  {https://ui.adsabs.harvard.edu/abs/2012MNRAS.427.1245R} {427, 1245}

\bibitem[\protect\citeauthoryear{{Ren}, {Chen}, {Zhang}, {de Grijs}, {Deng}  \&
  {Huang}}{{Ren} et~al.}{2021}]{Ren2021}
{Ren} F.,  {Chen} X.,  {Zhang} H.,  {de Grijs} R.,  {Deng} L.,   {Huang} Y.,
  2021, arXiv e-prints, \href
  {https://ui.adsabs.harvard.edu/abs/2021arXiv210316096R} {p. arXiv:2103.16096}

\bibitem[\protect\citeauthoryear{{Riello} et~al.,}{{Riello}
  et~al.}{2020}]{Riello2020}
{Riello} M.,  et~al., 2020, arXiv e-prints, \href
  {https://ui.adsabs.harvard.edu/abs/2020arXiv201201916R} {p. arXiv:2012.01916}

\bibitem[\protect\citeauthoryear{{Rodrigues} et~al.,}{{Rodrigues}
  et~al.}{2017}]{Rodrigues2017}
{Rodrigues} T.~S.,  et~al., 2017, \mn@doi [\mnras] {10.1093/mnras/stx120},
  \href {https://ui.adsabs.harvard.edu/abs/2017MNRAS.467.1433R} {467, 1433}

\bibitem[\protect\citeauthoryear{{Sharma}, {Stello}, {Bland-Hawthorn}, {Huber}
  \& {Bedding}}{{Sharma} et~al.}{2016}]{Sharma2016}
{Sharma} S.,  {Stello} D.,  {Bland-Hawthorn} J.,  {Huber} D.,   {Bedding}
  T.~R.,  2016, \mn@doi [\apj] {10.3847/0004-637X/822/1/15}, \href
  {https://ui.adsabs.harvard.edu/abs/2016ApJ...822...15S} {822, 15}

\bibitem[\protect\citeauthoryear{{Silva Aguirre} et~al.,}{{Silva Aguirre}
  et~al.}{2018}]{SilvaAguirre2018}
{Silva Aguirre} V.,  et~al., 2018, \mn@doi [\mnras] {10.1093/mnras/sty150},
  \href {https://ui.adsabs.harvard.edu/abs/2018MNRAS.475.5487S} {475, 5487}

\bibitem[\protect\citeauthoryear{{Skrutskie} et~al.,}{{Skrutskie}
  et~al.}{2006}]{Skrutskie2006}
{Skrutskie} M.~F.,  et~al., 2006, \mn@doi [\aj] {10.1086/498708}, \href
  {https://ui.adsabs.harvard.edu/abs/2006AJ....131.1163S} {131, 1163}

\bibitem[\protect\citeauthoryear{{Slumstrup}, {Grundahl}, {Silva Aguirre}  \&
  {Brogaard}}{{Slumstrup} et~al.}{2019}]{Slumstrup2019}
{Slumstrup} D.,  {Grundahl} F.,  {Silva Aguirre} V.,   {Brogaard} K.,  2019,
  \mn@doi [\aap] {10.1051/0004-6361/201833739}, \href
  {https://ui.adsabs.harvard.edu/abs/2019A&A...622A.111S} {622, A111}

\bibitem[\protect\citeauthoryear{{Stassun} \& {Torres}}{{Stassun} \&
  {Torres}}{2021}]{Stassun2021}
{Stassun} K.~G.,  {Torres} G.,  2021, \mn@doi [\apjl]
  {10.3847/2041-8213/abdaad}, \href
  {https://ui.adsabs.harvard.edu/abs/2021ApJ...907L..33S} {907, L33}

\bibitem[\protect\citeauthoryear{{Stello} et~al.,}{{Stello}
  et~al.}{2011}]{Stello2011}
{Stello} D.,  et~al., 2011, \mn@doi [\apj] {10.1088/0004-637X/739/1/13}, \href
  {https://ui.adsabs.harvard.edu/abs/2011ApJ...739...13S} {739, 13}

\bibitem[\protect\citeauthoryear{{Stetson}, {Bruntt}  \& {Grundahl}}{{Stetson}
  et~al.}{2003}]{Stetson2003}
{Stetson} P.~B.,  {Bruntt} H.,   {Grundahl} F.,  2003, \mn@doi [\pasp]
  {10.1086/368337}, \href
  {https://ui.adsabs.harvard.edu/abs/2003PASP..115..413S} {115, 413}

\bibitem[\protect\citeauthoryear{{Taylor}}{{Taylor}}{2005}]{Taylor2005}
{Taylor} M.~B.,  2005, in {Shopbell} P.,  {Britton} M.,   {Ebert} R.,  eds,
  Astronomical Society of the Pacific Conference Series Vol. 347, Astronomical
  Data Analysis Software and Systems XIV. p.~29

\bibitem[\protect\citeauthoryear{{Tofflemire}, {Gosnell}, {Mathieu}  \&
  {Platais}}{{Tofflemire} et~al.}{2014}]{Tofflemire2014}
{Tofflemire} B.~M.,  {Gosnell} N.~M.,  {Mathieu} R.~D.,   {Platais} I.,  2014,
  \mn@doi [\aj] {10.1088/0004-6256/148/4/61}, \href
  {https://ui.adsabs.harvard.edu/abs/2014AJ....148...61T} {148, 61}

\bibitem[\protect\citeauthoryear{{Ulrich}}{{Ulrich}}{1986}]{Ulrich1986}
{Ulrich} R.~K.,  1986, \mn@doi [\apjl] {10.1086/184700}, \href
  {https://ui.adsabs.harvard.edu/abs/1986ApJ...306L..37U} {306, L37}

\bibitem[\protect\citeauthoryear{{Vasiliev} \& {Baumgardt}}{{Vasiliev} \&
  {Baumgardt}}{2021}]{Vasiliev2021}
{Vasiliev} E.,  {Baumgardt} H.,  2021, arXiv e-prints, \href
  {https://ui.adsabs.harvard.edu/abs/2021arXiv210209568V} {p. arXiv:2102.09568}

\bibitem[\protect\citeauthoryear{{Villanova}, {Carraro}, {Geisler}, {Monaco}
  \& {Assmann}}{{Villanova} et~al.}{2018}]{Villanova2018}
{Villanova} S.,  {Carraro} G.,  {Geisler} D.,  {Monaco} L.,   {Assmann} P.,
  2018, \mn@doi [\apj] {10.3847/1538-4357/aae4e5}, \href
  {https://ui.adsabs.harvard.edu/abs/2018ApJ...867...34V} {867, 34}

\bibitem[\protect\citeauthoryear{{Warren} \& {Cole}}{{Warren} \&
  {Cole}}{2009}]{Warren2009}
{Warren} S.~R.,  {Cole} A.~A.,  2009, \mn@doi [\mnras]
  {10.1111/j.1365-2966.2008.14268.x}, \href
  {https://ui.adsabs.harvard.edu/abs/2009MNRAS.393..272W} {393, 272}

\bibitem[\protect\citeauthoryear{{Yong} et~al.,}{{Yong}
  et~al.}{2016}]{Yong2016}
{Yong} D.,  et~al., 2016, \mn@doi [\mnras] {10.1093/mnras/stw676}, \href
  {https://ui.adsabs.harvard.edu/abs/2016MNRAS.459..487Y} {459, 487}

\bibitem[\protect\citeauthoryear{{Zinn}}{{Zinn}}{2021}]{Zinn2021}
{Zinn} J.~C.,  2021, \mn@doi [\aj] {10.3847/1538-3881/abe936}, \href
  {https://ui.adsabs.harvard.edu/abs/2021AJ....161..214Z} {161, 214}

\makeatother
\end{thebibliography}








\bsp	
\label{lastpage}
\end{document}